\documentclass[tighten,twocolumn]{aastex631}

\usepackage[utf8]{inputenc}
\usepackage{amsmath}
\usepackage{graphicx}
\usepackage{float}
\usepackage{xfrac}

\newcommand{\RomanNumeralCaps}[1]{\MakeUppercase{\romannumeral #1}}

\received{XXXX}
\revised{XXXX}
\accepted{XXXX}
\submitjournal{AJ}

\shorttitle{Deconstructing Photospheric Spectral Lines in Solar and Stellar Flares}
\shortauthors{Monson et al.}

\graphicspath{{./}{}}

\begin{document}

\title{Deconstructing Photospheric Spectral Lines in Solar and Stellar Flares}

\correspondingauthor{Aaron Monson}
\email{amonson01@qub.ac.uk}

\author[0000-0002-3305-748X]{Aaron J. Monson}

\author{Mihalis Mathioudakis}
\affiliation{Astrophysics Research Centre, School of Mathematics and Physics, Queen’s University Belfast, BT7 1NN, Northern Ireland, UK}
 
 \author{Adam F. Kowalski}
\affiliation{National Solar Observatory, University of Colorado Boulder, 3665 Discovery Drive, Boulder, CO 80303, USA}
\affiliation{Department of Astrophysical and Planetary Sciences, University of Colorado Boulder, 2000 Colorado Avenue, Boulder, CO 80305, USA}
 \affiliation{Laboratory for Atmospheric and Space Physics, University of Colorado Boulder, 3665 Discovery Drive, Boulder, CO 80303, USA}

 \begin{abstract}

During solar flares, spectral lines formed in the photosphere have been shown to exhibit changes to their profiles despite the challenges of energy transfer to these depths. Recent work has shown that deep-forming spectral lines are subject to significant contributions from regions above the photosphere throughout the flaring period, resulting in a composite emergent intensity profile from multiple layers of the atmosphere. We employ radiative-hydrodynamic and radiative transfer calculations to simulate the response of the solar/stellar atmosphere to electron beam heating and synthesize spectral lines of Fe \RomanNumeralCaps{1} to investigate the line-of-sight velocity fields information available from Doppler shifts of the emergent intensity profile. By utilizing the contribution function to deconstruct the line profile shape into its constituent sources, we show that variations in the line profiles are primarily caused by changes in the chromosphere. Up-flows in this region were found to create blueshifts or "false" redshifts in the line core dependent on the relative contribution of the chromosphere compared to the photosphere. In extreme solar and stellar flare scenarios featuring explosive chromospheric condensations, red-shifted transient components can dominate the temporal evolution of the profile shape, requiring a tertiary component consideration to fully characterize. We conclude that deep-forming lines require a multi-component understanding and treatment, with different regions of the spectral line being useful for probing individual regions of the atmosphere's velocity flows.

 \end{abstract}


\section{Introduction} \label{sec:intro}

Solar and stellar flares comprise some of the most dynamic and impulsively generated events that stars endure during their main sequence lifetimes. Under the current widely accepted scenario, impulsive reconfiguration of the magnetic field lines in the outer solar atmosphere releases vast amounts of ``free'' energy. The magnetic energy released can be as high as 10$^{33}$ ergs and drives the heating of the surrounding plasma to tens of millions of degrees \citep{Fletcher11}. A similar framework is believed to be applicable to their stellar counterparts \citep{Haisch91}, but these can be orders of magnitude more energetic ($10^{34} - 10^{35}$ ergs) than their solar counterparts. 

Although the reconnection site is thought to be located in the corona, the majority of the energy released is deposited in the chromosphere \citep{Fletcher11,Milligan14}. There are numerous methods by which this energy is conveyed to the lower solar atmosphere including via thermal conduction \citep{Wamuth16,Wamuth20,Allred22}, Alfvén wave heating \citep{Fletcher08,Kerr16}, proton beams \citep{Prochazka18} and XEUV Backwarming \citep{Allred05,Lindsey08}. A widely accepted dominant energy transport mechanism during the impulsive phase, is via the acceleration of ambient electrons to near relativistic speeds \citep{Brown71,Emslie78,Holman11}. These ``beamed'' electrons lead to the energy transfer from where the energy is first released at coronal heights, to the chromosphere where they are thermalized via Coulomb collisions, heating the lower solar atmosphere to MK temperatures. Observational evidence for this heating process is the emission of Hard X-rays from the footpoints of flaring loop arcades by Bremmstrahlung radiation \citep{Hudson92,Neidig93}. \cite{Fisher85} showed that for heating rates $\leq 10^{10}$ erg cm$^{-2}$ s$^{-1}$ the chromosphere undergoes ``gentle'' evaporation, consisting of upwards motion of the plasma at tens km s$^{-1}$. For heating rates $> 3 \times 10^{10}$ erg cm$^{-2}$ s$^{-1}$ the chromosphere ``explosively'' evaporated, with upflows an order of magnitude faster than their gentle counterparts accompanied by dense momentum-conserving downflow condensations \citep{Canfield90,Milligan06}. The explosive evaporation energy threshold is not a strictly defined limit and is dependent upon the heating timescale, low-energy cutoff and power-law index of the beam \citep{Fisher89}. Such flows in the lower solar/stellar atmosphere induced by these electron beams have been extensively investigated both observationally and through radiative-hydrodynamic (RHD) modeling \citep{Milligan09,Kuridze15,Graham15,Kowalski17a,Kowalski17,Graham20}. Asymmetric line profiles and Doppler shifts are the primary methods to infer the changes in the physical conditions of the solar \& stellar atmospheres' reaction to flare heating. Using high-cadence observations from the Interface Region Imaging Spectrograph; IRIS \citep{DePontieu14} of the impulsive phase of solar flares \cite{Graham20} showed clear evidence for multiple components on several chromospheric transitions of Fe \RomanNumeralCaps{1} \& \RomanNumeralCaps{2}. These were interpreted as an enhanced stationary component and a red-shifted ``satellite'' component. Insights gained through the radiative-hydrodynamic code RADYN revealed that the origin of these spectral features is due to the dense downflowing chromospheric condensations. This agrees with the description of \cite{Kowalski17a} pertaining to a two-layer emission, comprising of both stationary and downwards-moving components.

While chromospheric lines have been extensively investigated, spectroscopic studies of deep-forming photospheric lines in flares is lacking. \cite{Sharykin17} showed that enhancements in the 617.3 nm Fe \RomanNumeralCaps{1} spectral line observed by the Solar Dynamics Observatory's Helioseismic and Magnetic Imager (SDO/HMI, \cite{Pesnell12, Scherrer12}) during the X-class flare of 2012 October 23 were co-spatial with the hard X-ray footpoints, indicative of the sites of electron beam heating, but delayed by several seconds. They also noted a redshift of the spectral line profile of around 0.5 km s$^{-1}$ during the period of beam heating. \cite{Hong18} modeled the response of the same spectral line in quiet-sun and penumbra atmospheres, finding that the cooler penumbra resulted in greater heating. \cite{Monson21}, hereafter referred as Paper 1, showed that spectral line core enhancements in three spectral lines of Fe \RomanNumeralCaps{1} are not due to increased photospheric emission from heating, but instead they are due to significant emission contribution from the chromosphere at these wavelengths. These resulted in deep-forming line profiles being sensitive to chromospheric velocities, resulting in false redshifts which would incorrectly indicate downflows in the photosphere.

Investigating how these deep-forming spectral line profiles are modified during solar \& stellar flares is crucial for understanding the energy deposition processes. The purpose of this work is to investigate the potential retrievable information available from photospheric Fe \RomanNumeralCaps{1} lines. Exploring the emission profile shapes these lines exhibit throughout the temporal evolution of several radiative-hydrodynamic simulations of flare scenarios and the physical atmospheric conditions which drive features. This paper will be structured as follows. Section \ref{sec:RHD} describes the radiative-hydrodynamic modeling tools utilized in this study, Section \ref{sec:Models} discusses the three flare scenarios for which spectral line profiles are synthesized for and discussed in Section \ref{sec:LineProfiles}. The physical processes which lead to the emergent profile shapes, specifically the atmospheric conditions which drive changes in the profile shape are discussed in Section \ref{sec:Atmospheric_Condtions}. Section \ref{sec:Line_Decon} then analyzes how deconstructing the flaring emission spectrum into their constituent parts can provide diagnostics of atmospheric mass flows, comparing these to the insights we gained from Section \ref{sec:Atmospheric_Condtions}. In Section \ref{sec:Conclusions} we summarize the usefulness of the deconvolution of complex flaring profiles and the applicability of the results these models provide to potential observables. 

\section{Radiative-Hydrodynamic Modeling}
\label{sec:RHD}

The response of the atmosphere to the energy injection in the form of a power-law distribution of electrons is first modeled using the RADYN RHD code \citep{Carlsson92,Carlsson95,Carlsson97,Allred15}. RADYN solves the coupled equations of hydrodynamics, charge conservation, atomic level populations, and radiative transfer rates on an adaptive grid \citep{Dorfi87}. The adaptive grid is weighted to resolve key features in the model atmospheres. In flares, such features are caused by steep temperature, density, and velocity gradients, and it is crucial that they are resolved for the accurate calculation of the radiative transfer. The electron beam heating distribution throughout the atmosphere is calculated using the Fokker-Planck equations for non-thermal particle beam transport and thermalisation \citep{McTiernan90}, for which no assumption of the target temperature is necessary, implemented by \citep{Allred15}. RADYN uses a six-level hydrogen atom, a nine-level helium, and a six-level Ca \RomanNumeralCaps{2} ion all with continuum. This allows the calculation of a number of transitions that are important for the chromospheric energy balance in NLTE \citep{Allred15}. RADYN allows for a range of atmospheric structures and electron beam parameters to be utilized for bespoke models of specific events \citep{Rubio16, Kuridze16, Graham20, Kowalski22ADLeo} or the creation of a large grid of electron beam parameters such as the F-CHROMA project's \citep{Carlsson23}.

As the Fe \RomanNumeralCaps{1} lines of interest in this study are not included in RADYN, and need to be treated in NLTE \citep{Holzreuter12,Holzreuter13,Holzreuter15,Smitha20,Smitha21,Smitha22}, it becomes necessary to utilize an additional radiative transfer code. The RH code \citep{Uitenbroek01} can take snapshots from a RADYN model as input to perform the radiative transfer and spectral synthesis of the Fe lines. RH is a time-independent code, thereby losing the model atmosphere's non-equilibrium (NE) atomic level population history for each timestep. It instead uses the NE electron densities from RADYN to perform radiative transfer calculations in statistical equilibrium. The combination of RADYN \& RH allows us to gain insight into the physical conditions in the solar atmosphere which drive changes in the emergent intensities of spectral lines and continuum. 

\section{Solar \& Stellar Flare Models} \label{sec:Models}

\begin{figure}[!hb]
    \centering
    \includegraphics[width=\linewidth]{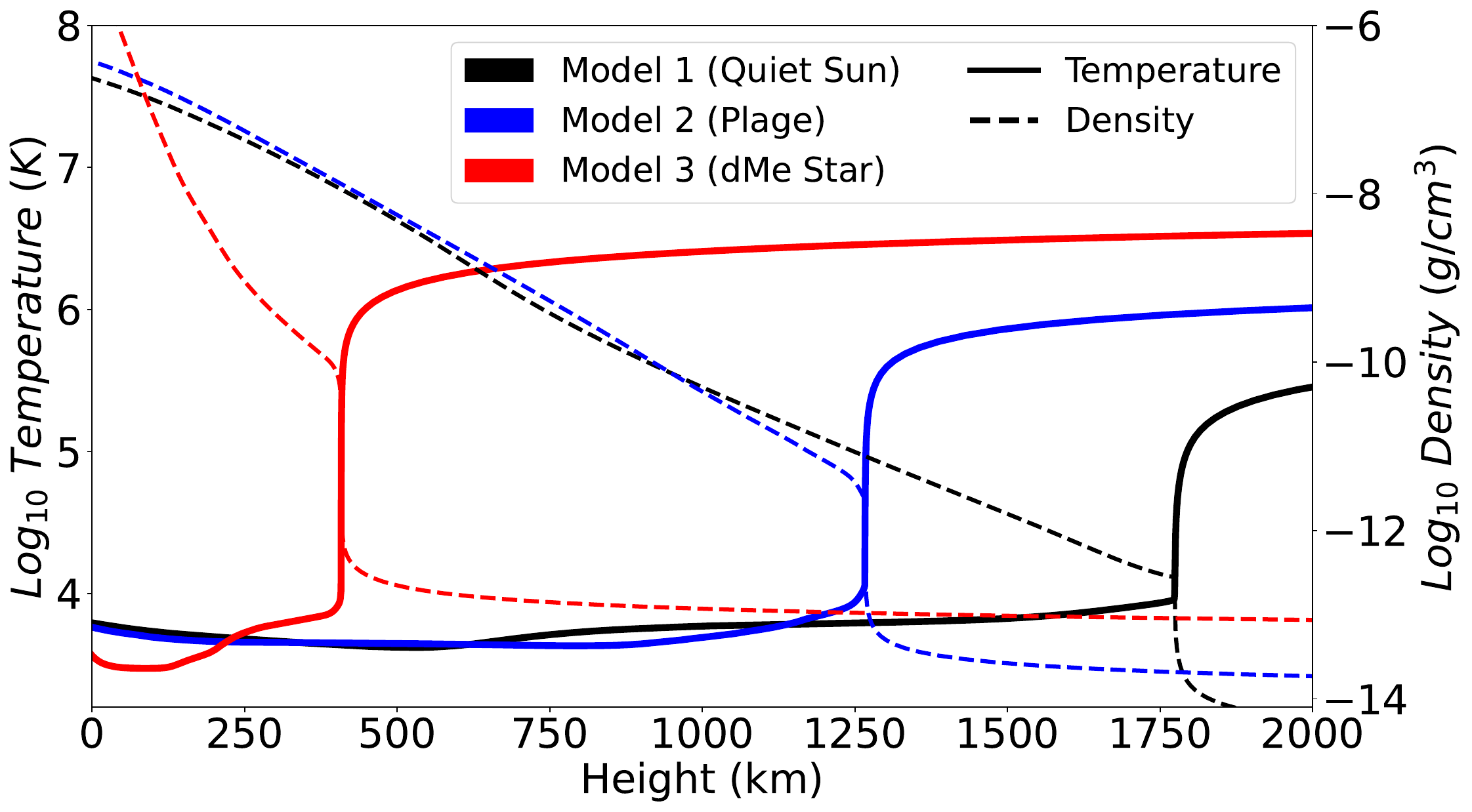}
    \caption{The starting atmospheric conditions for the three RADYN models utilized in this study. An FCHROMA F12 model (black) with a TR at $\sim$1750 km, the model utilized by \cite{Zhu19} (blue) with a lower TR ($\sim$1250 km) and slightly cooler chromosphere, and an M dwarf atmosphere (red). Solid lines show the temperature structure of each model from the photospheric floor (where $\tau_{500nm} =1)$ to heights above the transition region. Overplotted in dashed lines is the density structure of each model.}
    \label{fig:atomos_profiles}
\end{figure}

This work utilizes three different atmospheric stratifications, each heated by an accelerated electron beam. The initial (before any beam heating) temperature and density structure of the three models are shown in Figure \ref{fig:atomos_profiles}. All three atmospheres take the form of a quarter-circle loop with a length of 10 Mm extending from the sub-photosphere to the corona. In each model a height of 0 corresponds to the $\tau_{500nm} =1$ layer in the atmosphere, a standard definition for the photosphere \citep{Norton06,Hong18,Svanda18}. For the purpose of this work, we utilize the zero height definition as our photosphere lower limit and define the temperature minimum region (TMR) as the upper boundary of the photosphere.

Model 1 (black lines in Figure \ref{fig:atomos_profiles}) uses a pre-flare atmosphere similar to the VAL3C semi-emprical model of \cite{Vernazza81}, extended with a transition region and corona, generated as part of the F-CHROMA project \citep{Carlsson23}. This model has the highest transition region (TR) of the three models, a photosphere upper boundary of 550 km, and a photosphere to TR distance of 1200 km. Model 2 (blue lines) is a solar plage model utilized and detailed in \cite{Zhu19}. This model features a cooler chromosphere with a lower TR compared to Model 1, but similar density and temperature structures below $\sim600$ km. Notably, the effect of this cooler plage atmosphere is a very high TMR compared to Model 1 ($\sim 800$ km) resulting in a larger effective photosphere as per our definition of the upper boundary. The step down of the TR to lower geometric heights in Model 2 provides a test ground for a smaller photosphere-to-TR height gap ($\sim 470$ km), which has a significant effect on the distribution of the electron beam heating through the atmosphere as discussed in Section \ref{sec:Atmospheric_Condtions}. 

\begin{figure*}[]
    \centering
    \includegraphics[width=\linewidth]{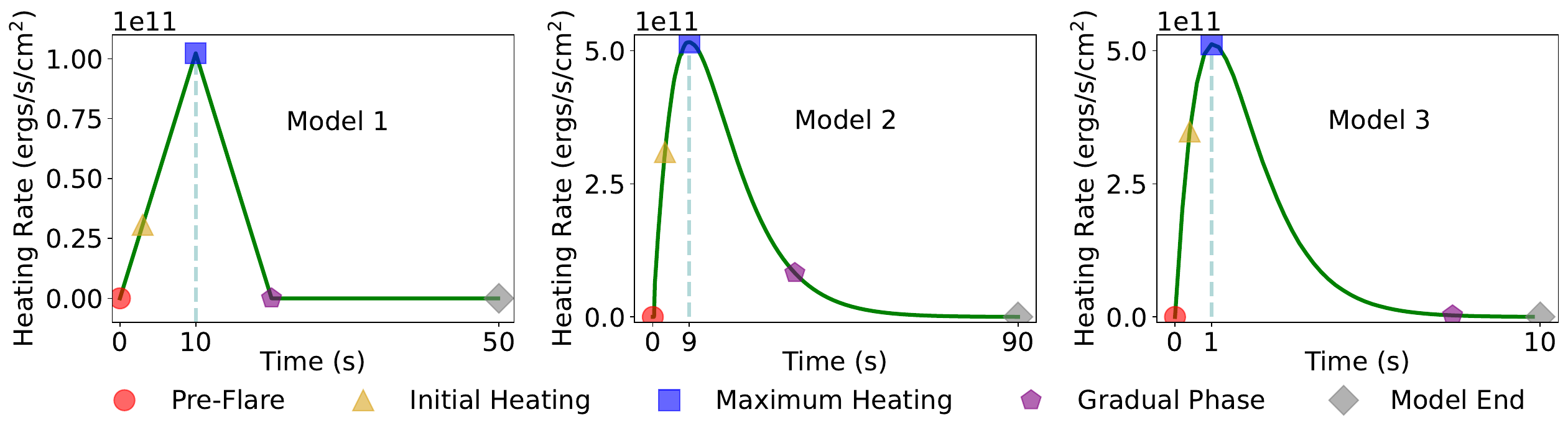}
    \caption{The temporal evolution of the three beam heating profiles for the atmospheric structures utilized in this study. Here the Heating Rate refers to the total energy flux density injected into the model atmosphere at a given time. Key parameters of the beam composition and timings are summarized in Table \ref{tab:EBP}. Markers for the times discussed in Figure \ref{fig:Line_profiles} are overplotted for each model for context.}
    \label{fig:Heating_profiles}
\end{figure*}

Finally, Model 3 details a stellar flare on the M dwarf star AD Leo (dM3e classification) \citep{Kowalski22ADLeo}, with a higher effective surface gravity, $\log(g) = 4.75$ as opposed to 4.44 in the solar models, leading to a comparatively far more compact atmosphere. The rapidly rising density profile shown in Figure \ref{fig:atomos_profiles} (red dashed lines) compared to the density profiles of the solar models, along with the relatively short TR-to-TMR range ($\sim$300 km) provide the conditions for significant heating of the deep chromosphere by the electron beams, which is difficult to achieve in solar flares \citep{Kowalski23}. The three model atmospheres are heated by the electron beam profiles described in Table \ref{tab:EBP}. All models utilize the same spectral index ($\delta$ = \! 4) and low energy cutoff ($E_{c}=25$ keV), but differing amounts of total energy and energy rates. The beam heating profiles for each model are also shown in Figure \ref{fig:Heating_profiles}.

\begin{deluxetable*}{cccccccc}[]
\tabletypesize{\scriptsize}
\tablewidth{0pt}
\tablecaption{The key properties of the models utilized in this study \label{tab:EBP}}

\tablehead{\colhead{Model} & \colhead{Model Duration} & \colhead{Heating Profile Shape} & \colhead{Total Energy} & \colhead{Peak Beam Time}  & \colhead{t$_{1/2}$}  & \colhead{Spectral Index}  & \colhead{Low Energy Cutoff} \\ 
\colhead{} & \colhead{(s)} & \colhead{} & \colhead{(ergs cm$^{-2)}$} & \colhead{(s)}  & \colhead{(s)} & \colhead{($\delta$)}  & \colhead{(keV)} }
\startdata
Model 1 \citep{Carlsson23} & 50  & Triangular & $1 \times \ 10^{12}$    &  10 & 10  & 4  & 25  \\ 
Model 2 \citep{Zhu19}   & 90  & Pulsed       & $1.1 \times \ 10^{13}$  &  9  & 20  & 4  & 25  \\ 
Model 3 \citep{Kowalski22ADLeo}   & 9.8 & Pulsed       & $1.28 \times \ 10^{12}$ &  1  & 2.3 & 4  & 25 \\ 
\enddata

\end{deluxetable*}

Model 1 utilizes a triangular heating profile over 20 s giving a total of $10^{12}$ erg cm$^{-2}$ (F12), followed by 30s where the atmosphere evolves without heating. Models 2 \& 3 utilize the pulsed energy injection description of \cite{Aschwanden04}, each following unique timescales to their heating times and their total energies summarized in Table \ref{tab:EBP}. To easily compare model heating times and energy deposition rates we utilize the t$_{1/2}$ time definition employed by \cite{Kowalski13}, the timing between the two moments where the energy deposition is half that of the peak energy deposition. This timescale helps us contextualize the amount of energy deposited over the lifetime of the heating pulse. Model 3 utilizes a shorter heating timescale compared to the two solar flare cases to simulate the \textit{``impulsive flares''} of \cite{Kowalski13,Kowalski22ADLeo}. The very short ($<10$ s) timescales for nonthermal heating are chosen to conform with the very short rise times observed for M dwarf flare UV, optical, and U-band rise times \citep{Mathioudakis,MacGregor21,Kowalski16}.

The aim of this paper is to explore how these differing atmospheres and unique heating profiles result in variations in the emergent spectral line profile shapes. The key features of the line profiles can be interpreted using the physical parameters of the atmosphere known from the RADYN inputs. This interpretation is aided by multi-component contribution function analysis of the complex profile shapes that arise from the heating process.

\section{Atmospheric evolution and associated contribution functions} \label{sec:Atmospheric_Condtions}

\begin{figure}[!h]
    \centering
    \includegraphics[width=\linewidth]{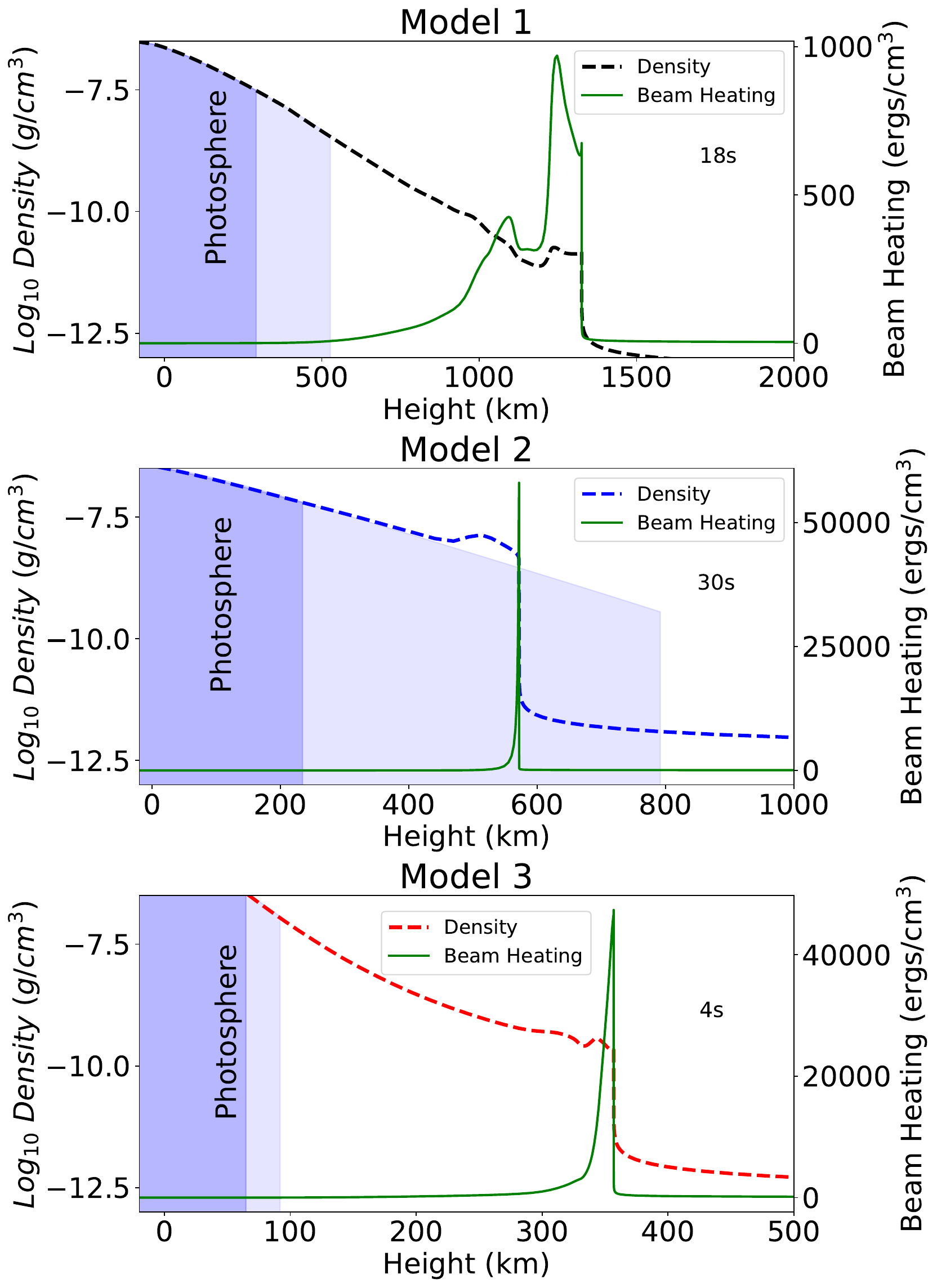}
    \caption{Snapshots from the three models utilized in this study for 3 times after the peak heating rate. The mass density for each model (dashed line) is also shown together with the volumetric electron beam heating rate (green solid line) corresponding to the right axis. Photosphere-classified height ranges for each model at the stated time are shown as blue-shaded regions, with the lighter-shaded region contextualizing the t=0s pre-flare photosphere.}
    \label{fig:Density_Bheat}
\end{figure}

RHD modeling allows unique insights into the physical mechanisms within the solar or stellar atmosphere which drive the changes to the spectral line profiles we observe. Firstly, understanding how the electron beam heating is distributed throughout the atmosphere, and the secondary processes induced (e.g. Backwarming, Thermal Conduction), is vital for detailing how Doppler motions are formed to create dynamic spectral line profiles. Figure \ref{fig:Density_Bheat} shows the heating distribution throughout each of the model atmospheres during the decay phase of each model. The dark blue shaded region shows the photospheric height range at the stated time, with the lighter blue shaded region indicating the $t=0s$ photosphere. In all cases, the TMR is pushed deeper into the atmosphere as the chromosphere expands, dropping by several hundred kilometers in the two solar models. Model 2 shows the largest reduction of the photospheric ceiling.

The location and extent of the electron beam heating (green solid line) differ between the three models. Model 1 (Top Panel of Figure \ref{fig:Density_Bheat}) shows the broadest heating range, with significant direct beam heating over $\sim500$ km. The primary heating maximum, in this case, is shown to be co-spatial with an increase in density at 1300 km and is formed by material that is swept into the lower atmosphere by dramatic heating just below the transition region. A secondary peak in the heating is observed lower in the atmosphere below this condensation feature when the largely undisturbed atmospheric mass density exceeds $\sim 10^{-11}$ g cm$^{-3}$. Below this height, the beam heating decays with depth, resulting in non-significant rates of direct heating before reaching the photosphere. This profile is in good agreement with the \cite{Kowalski17} framework whereby the deeper secondary heating peak is a result of higher energy electrons in the beam which can penetrate through the density structure formed by lower energy electrons at greater atmospheric heights. 

Models 2 \& 3 show far more localized regions with 10s km scale slabs of dense material being heated by an order of magnitude more direct beam heating ($ >10^4$ erg cm$^{-3}$) compared to Model 1. In both cases, the heating is located around a local maximum in the density (lower panels Figure \ref{fig:Density_Bheat}). The condensation in Model 2 is formed from just below the transition region ($t=0$ s; $1250$ km) and propagates through the atmosphere, dissipating into the lower chromosphere/upper photosphere after 60s at a height of $\sim$520 km. During the later stages of the model, in the final 30s once the beam heating rate has diminished, the transition region and transferred mass from the condensation begins to restore upwards. The atmosphere does not fully restore by the end time of the model.
 
Model 3 shows a similar localized electron beam heating formed around the condensation point, but over a broader range ($\sim$50 km) compared to Model 2. Due to the much longer timescale of the evolved state of Model 2 shown compared to Model 3 (30s evolution compared to 4 s) the condensation front in Model 3 has permeated significantly deeper into the atmosphere, reaching lower-chromospheric densities ($\sim 10^{-8}$ g cm$^{-3}$). The perceived jump in the mass density through the compression results in an effective barrier preventing the further downward propagation of the beam, leading to the very short height scale of energy deposition and the extreme peak in the heating. Compared to Model 2, the mass density jump through the condensation in Model 3 is much smaller, jumping just over 2 orders of magnitude compared to the near 4 orders jump in density in Model 2. This combined with the overall lower densities below the condensation in Model 3 (See Figure \ref{fig:atomos_profiles} Model 2 $\sim 600$ km \& Model 3 $\sim 350$ km comparative densities) leads to a broader beam heating peak compared to Model 2's localized heating.

Chromospheric condensations can have significant effects on the emission spectra of flaring atmospheres, creating red asymmetries, satellite components, and opacity effects \citep{Kuridze16, Kowalski17, Brown18, Graham20}. These effects can be studied further using the line contribution function \cite{Carlsson95} which is defined as:

\begin{equation}\label{eq:contrib_function}
\centering
 I_\nu = \int_{z_{\small0}}^{z_{\small1}} C_{\nu} \,dz =  \int_{z_{\small0}}^{z_{\small1}}S_{\nu} \, \tau_{\nu} \, e^{-\tau_{\nu}} \, \frac{\chi_{\nu}}{\tau_{\nu}} \,dz
\end{equation}

Where the emergent intensity at a specific frequency, \textit{I$_\nu$}, is defined as the integral of the contribution function, C$_\nu$, over a given height range. The contribution function is the product of the Source function, S$_\nu$, optical depth, $\tau_\nu$, and density of emitting material, $\chi_\nu$ each at a given frequency.

\begin{figure}[!h]
    \centering
    \includegraphics[width=\linewidth]{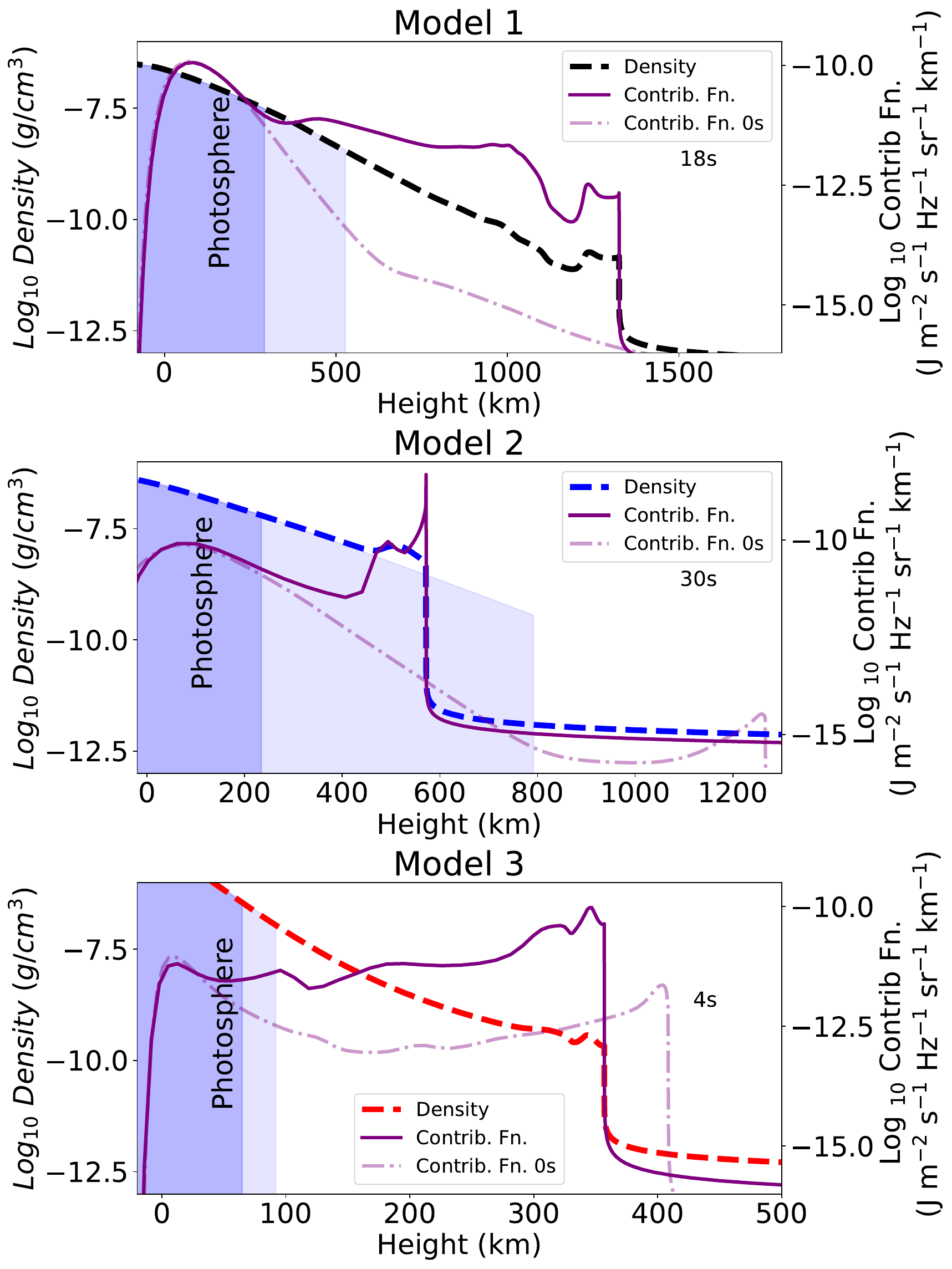}
    \caption{The same mass density structure for each model's evolved state as shown in Figure \ref{fig:Density_Bheat} over the height range of the lower solar atmosphere (dashed colored lines in each panel). Overplotted is the contribution function (Equation \ref{eq:contrib_function}) for the rest wavelength of the 630.1 nm Fe \RomanNumeralCaps{1} line at the pre-flare (dahsed-dot line) and evolved state (solid purple line) time stated in each panel. Blue shaded regions have the same meanings as Figure \ref{fig:Density_Bheat}.}
    \label{fig:Density_Contrib}
\end{figure}

Figure \ref{fig:Density_Contrib} shows the same density structure and as in Figure \ref{fig:Density_Bheat} with the contribution function for the 630.1 nm Fe \RomanNumeralCaps{1} line at the stated times (solid purple line). The pre-flare contribution function is also shown as a dashed-dotted line. The 630.1 nm line is predominantly photospheric, with a maximum contribution of $\sim 10^{-10}$ J m$^{-2}$ s$^{-1}$ Hz$^{-1}$ sr$^{-1}$ km$^{-1}$ in the two solar models and an order of magnitude lower, though still the clear maximum, formed around 40 km in Model 3. Both Models 1 \& 2 show significant decreases in the contribution function with increasing height, signifying the minimal amounts upper regions (the chromosphere and corona) contribute to these lines in the pre-flare atmosphere. The slower decline of the pre-flare contribution with increasing height in Model 2, along with the broader photospheric peak compared to Model 1, are a result of the cooler plage atmosphere. The significantly longer height range to the TMR extends the contribution function's tail-off due to more favorable formation temperatures for the transition. Model 3 shows a more extended, with a semi-constant contribution function through the chromosphere until a sudden drop upon hitting the transition region at 400 km. Despite this elongated structure to the contribution function, the plateau above the photosphere only contributes a few percent to the spectral line's emergent intensity. 

Models 2 \& 3 also show a secondary peak to their pre-flare contribution functions several kilometers below their transition regions due to an increased electron density in the hydrostatic equilibrium pre-flare atmosphere. In Model 2 this is only a small increase shown by the uptick in the dash-dot line just above 1200 km and does not significantly contribute to the total emergent intensity. For Model 3, the contribution function increase at the base of the transition region at $\sim 400$ km is more comparable to the photospheric peak contribution function values and contributes a few percent to the pre-flare line core intensity.

The effect of the electron beam heating, and the secondary heating processes induced from it, is immediately evident from the change in the contribution functions for all three models' evolved states (solid purple lines). Peaks in the contribution functions around the regions of density increase are primarily driven by the electron beam heating directly. Increases in the contribution function below the condensation in the lower chromosphere/upper photosphere are induced by backwarming heating, mostly by absorption of Balmer continuum radiation. All three models exhibit minimal changes in their contribution functions at depths where their lower maxima are located. The evolved state contribution of Model 1 shows significant emission from the lower chromosphere ($\sim 500-1000$ km range), and a secondary peak co-spatial with an increase in the mass density. While the line core remains dominated by the photosphere, approximately 20\% of the emergent intensity is emitted above the photosphere in agreement with the conclusions of Paper 1. The enhancement in the chromospheric contribution is temporally close to the evolution of the beam heating profile (Figure \ref{fig:Heating_profiles}). The contribution from regions about the photosphere reach a maximum within several seconds of the maximum beam heating before diminishing and restoring to the pre-flare contribution function profile. 

Model 2 (middle panel of Figure \ref{fig:Density_Contrib}) shows significantly more emission from the region of the condensation, with an order of magnitude increase in the contribution around the 400-600 km range. The contribution from this region exceeds that of the photospheric emission, resulting in a chromosphere-dominated spectral line profile. In particular, a thin ($<1$ km) slab of material at the back of the condensation, bordering the new transition region height, shows an additional order of magnitude increase in the contribution function spiking at $10^{-8.5}$ J m$^{-2}$ s$^{-1}$ Hz$^{-1}$ sr$^{-1}$ km$^{-1}$. Despite the 200 km width of the condensation's increased contribution, one-third of the feature's contribution to the 630.1 nm line core comes from this kilometre-broad region alone. Outside the region that shows the mass density increase, the contribution function profile closely follows changes to the density profile, indicative of the importance of the density of emitters, $\chi_\nu$, along with the MK temperatures above the condensation being unfavorable for Fe \RomanNumeralCaps{1} formation. Unlike Model 1, the emission from the chromosphere and condensation does not drop off with the reduction of the beam heating rate. The lower chromosphere between the photosphere and the condensation, $\sim250-450$ km range in Figure \ref{fig:Density_Contrib} middle panel, continues to contribute an order of magnitude more than the pre-flare state, while the emission from the condensation also continues to exceed this region. The effect of the condensation's movement through the lower solar atmosphere is evident by an increase in the contribution function throughout the entire model, tracking the feature's downwards motion and slowing down with the increased densities, even highlighting the upwards restoration of the atmosphere during the final stages in the model. 

Model 3 shows a similar profile and temporal evolution of the contribution function as Model 2, with a $<10^2$ increase in the plateaued contribution function in the chromosphere. The major rise in the contribution function coincides with the sudden rise in density in the condensation. The increase in contribution function in the lower chromosphere (the region between the photosphere and condensation) is so significant in this case that it exceeds the photospheric peak contribution, resulting in $<80\%$ of the line core emission originating from the lower chromosphere. This differs from Model 2, where the condensation was the primary region to cause an enhancement to the line core intensities. As the condensation sweeps lower into the atmosphere during the late stages of the model (8-10s), the contribution from this region becomes considerably more relevant and is discussed in Section \ref{sec:Line_Decon}.

Understanding the evolution of the flaring atmosphere is vital for reconstructing the emerging complex line profiles. The pre-flare photospheric line profiles can become chromospherically dominated during flares (Models 2 and 3) depending on the depth reached by the condensation and the energetics of the event. With this understanding we now study the emergent line profiles from these models to discern if the changes to the atmospheric structure are reflected in the photospheric spectral lines.

\section{Emergent Line Profiles}\label{sec:LineProfiles}

The atomic data file used to create these synthetic line profiles consists of a 31-level Fe I atom, with a singular once-ionized Fe II state and a double-ionized Fe III state. 63 bound-bound transitions, 31 bound-free transitions, and one fixed transition are detailed in the atom file, including collisional excitation and ionization rates to the two ionized states. As Model 3 represents an M dwarf star, the relative abundance of iron was adjusted to [Fe/H]$= 0.28 \pm 0.17$ following the value of AD Leo reported by \cite{Rojas12}.

\begin{figure*}[!h]
    \centering
    \includegraphics[width=\linewidth]{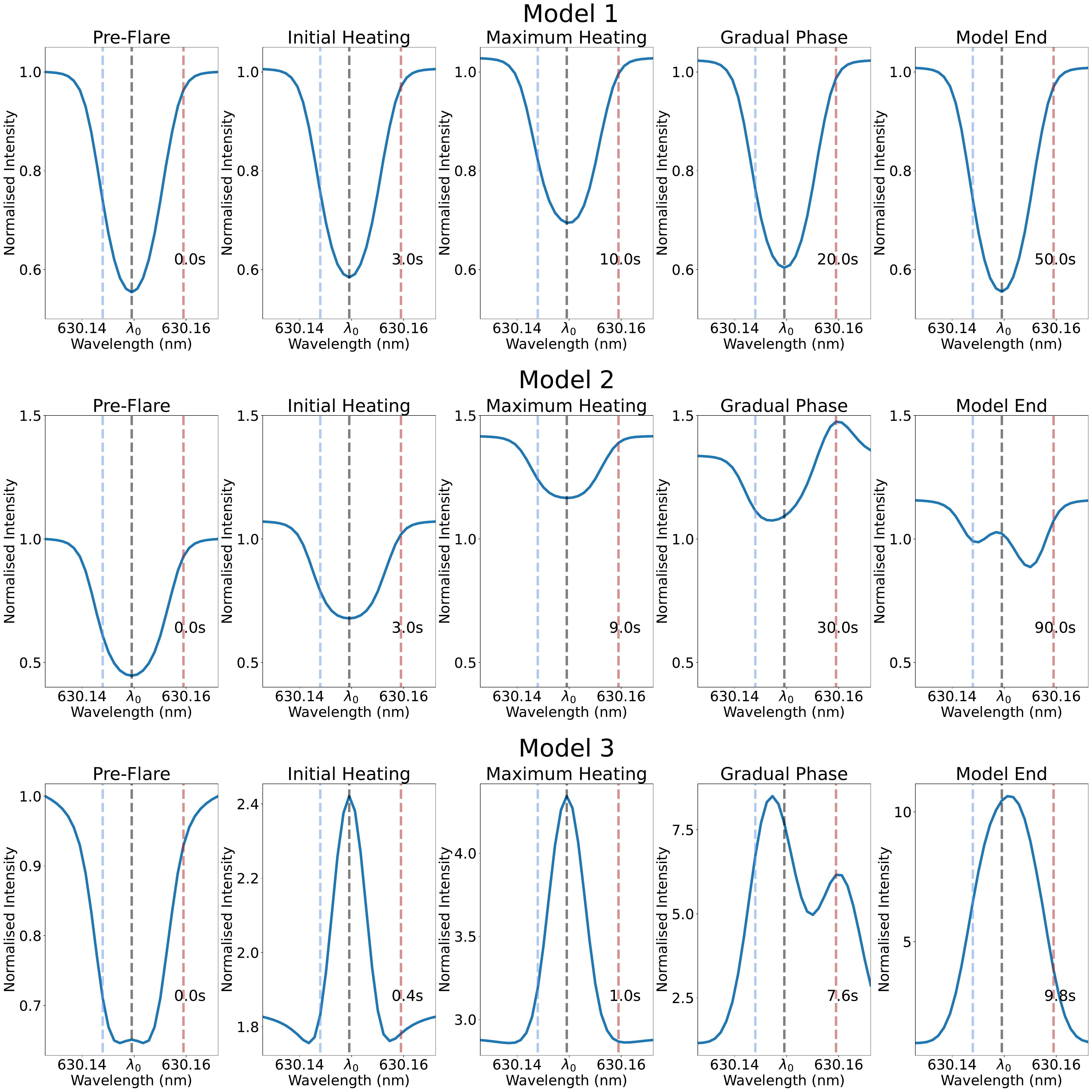}
    \caption{The temporal evolution of the 630.1 nm Fe \RomanNumeralCaps{1} spectral line profile for the 3 models. Top row; The quiet sun pre-flare atmosphere (Model 1). Middle row; Plage pre-flare atmosphere (Model 2). Bottom row; dMe star stellar flare (Model 3). ``Initial Heating'' times for each model are taken from when 5\% of the beam's energy has been deposited. ``Gradual Phase'' times are selected to show the transiting features in the line profile between peak beam heating and the end of the model. All intensities are normalized to a model's pre-flare continuum level.}
    \label{fig:Line_profiles}
\end{figure*}

The synthesized line profiles for the 630.1 nm Fe \RomanNumeralCaps{1} in each of the above-described models are shown in Figure \ref{fig:Line_profiles} for different times throughout the simulation. As discussed in Paper 1, the 630.1 nm line shows the most dynamic effects and exhibits the highest chromospheric contribution of the three Fe \RomanNumeralCaps{1} lines analyzed here and shall be the focus of this work. A summary of the 617.3 nm and 630.2 nm lines can be found in Appendix A. Each profile in Figure \ref{fig:Line_profiles} is normalized to the pre-flare continuum intensity to highlight better the enhanced features' scale. Additionally, Models 1 \& 2 have fixed intensity scales at all stages to easily contextualize the profile's enhancement compared to the pre-flare state. This was not possible for Model 3 due to the extent of the enhancement.

As the 3 models have different heating rates and heating profiles, we show the spectral line profiles at 5 stages of key interest to allow comparison. The times of each of these stages are overplotted as shape symbols in Figure \ref{fig:Heating_profiles}. The stages are defined as; the ``Pre-Flare'' profile before beam heating occurs (red circle), an ``Initial Heating'' at the time when 5\% of a model's total beam energy has been deposited (yellow triangle), and the time of ``Maximum Heating'' when the electron beam flux reaches its maximum ergs cm$^{-3}$ value (blue square). The ``Gradual Phase'' time (purple pentagon) does not have a consistent physical event like earlier defined stages, but instead is a selected time of importance to contextualize the atmosphere and line profile's evolution between the Maximum Heating time and the latest time available in the model atmosphere's evolution (Model End; gray diamond). These five categories provide insight into the key snapshots of the profiles' evolution which can then be matched to the physical conditions in the RADYN atmospheres.

In Figure \ref{fig:Line_profiles}, Model 1 shows an increase in the line core intensity, similar to the results in Paper 1, reaching a maximum several seconds after the maximum beam heating. This is in agreement with \cite{Sharykin17}, but with a greater enhancement, ($\sim$25\%) compared to the pre-flare than the 3F11 models utilized in Paper 1 ($\sim$10\% enhancement). The temporal evolution of the line closely follows the beam heating rate, with the line core depth decreasing during the stages of initial heating and up to the Maximum Heating. Following this time ($t=10s$), the line reaches its greatest enhancement ($t=12s$) before its intensity begins to diminish (Gradual Phase panel of Figure \ref{fig:Line_profiles} Model 1). This provides the simplest profile evolution of the three models and highlights that a three-fold increase in the electron beam's total energy alone does not drive the complex variations in profile shape in a solar-type atmosphere discussed next.

Model 2 is exposed to a ten times more energetic beam compared to Model 1 and features a lower TR height with a cooler chromosphere as discussed in Section \ref{sec:Models}. The pre-flare line profile for this model has a deeper line core than Model 1, but a far more significant increase in core intensity during the initial heating ($\sim$50\% filling; $0.3$ normalized intensity continuum-core difference compared to the pre-flare $0.55$). Between this time and the Beam Maximum, the entire continuum and profile are enhanced to exceed pre-flare values (Row 2, middle panel all values for normalized intensity exceed 1). Consistent with Model 1, Paper 1, and \cite{Sharykin17}, the maximum enhancement to the line core is reached within several seconds of the maximum electron beam heating rate also showing the shallowest continuum-core intensity difference. At later stages of the model, a transient asymmetry emerges in the red wing of the line profile (Model 2 Gradual Phase panel). This feature migrates through the line profile for the following 20s, creating a ``W-shape'' central reversal, before forming a blue side enhancement in the Model End panel of Figure \ref{fig:Line_profiles}. This transient Doppler-shifted emission feature is a clear difference between the Model 1 line profiles or those discussed in Paper 1, showing the contribution from the chromospheric condensation's effect on the total emergent profile further discussed in Section \ref{sec:Line_Decon}.

The M Dwarf stellar flare case (Model 3) is also detailed in the bottom row of Figure \ref{fig:Line_profiles}. The pre-flare profiles of the three Fe I lines show varying degrees of a central reversal to their line cores, which were not present in the solar models, due to the increased contribution from the base of the transition region discussed previously for Figure \ref{fig:Density_Contrib}. The effect of the much shorter heating timescale is evident by the immediate reversal of the pre-flare absorption profile into emission by the Initial Heating time ($t=0.4s$) along with significant continuum enhancement bringing the entire profile well above the pre-flare level. The single emission peak continues to increase in intensity through the next several seconds, developing to the profile seen by the Maximum Heating time panel in the bottom row with the line core exceeding four times the pre-flare near-continuum values. The depth of the line profile is significantly greater by this time compared to the Initial Heating stage, indicating that wavelengths closer to the rest wavelength experience a significant increase in emission contribution beyond just the continuum enhancement. The line core continues to increase in intensity for another second after the time of maximum beam heating, reaching an initial intensity maximum five times the pre-flare continuum values while the continuum remains similar to the values at the Maximum Heating stage. This second delay between the time of maximum beam heating and the maximum core enhancement is consistent with the profile evolutions seen for Models 1 \& 2 and in Paper 1, though this is a slightly shorter delay than in these cases.

For the following 2 seconds after intensity maximum, the entire profile shows an initial uniform decrease in intensity, a similar temporal evolution as other models where the intensity enhancement is temporally close to the beam heating rate. During this period, the line profile remains in emission and shows an increasing blueshift up to $\sim -1$ km s$^{-1}$ indicating upflows. While the near-continuum at shorter wavelengths continues to reduce in intensity as the beam heating diminishes, nearly returning to a normalized intensity of 1 by the Gradual Phase of Model 3, the blue-shifted emission peak shows a secondary increasing enhancement and exceeds the previous peak core intensity by 6 s into the model. This forms the brighter of the bimodal peaks shown in Model 3's Gradual Phase, with a far greater peak intensity and greater line depth than seen previously. Additionally, during this period, a significant redshifted emission peak emerges at longer wavelengths creating the second, smaller, redshifted emission peak in the Gradual Phase panel. The bimodal structure shown in this panel clearly indicates a developed velocity field consisting of both upflows and downflows that were not present at earlier times. As in Model 2, the transient redshifted component continues to migrate into the line core for the remaining 2 seconds of the model, resulting in the singular, broad, and brightest emission peak shown in the Model End panel. This singular peak exceeds 10 times the pre-flare continuum intensity and shows a clear redshifted component. The near-continuum wavelengths of the line have restored very close to their pre-flare intensities, showing 1.1 times the pre-flare continuum values despite the rest of the line profile clearly being far from restoring by the end of the model.

It is clear from the changes in the line profiles for Models 2 \& 3 that these photospheric lines are subject to complex, multi-components, formation effects leading to the dynamic evolution shown in Figure \ref{fig:Line_profiles}. The physical characteristics which result in these profiles must be understood in order to infer the meaningful retrievable information gained by observations of these spectral lines during flares.

\section{Spectral Line Contirbution Function Deconstruction}\label{sec:Line_Decon}

\begin{figure*}[!ht]
    \centering
    \includegraphics[width=\linewidth]{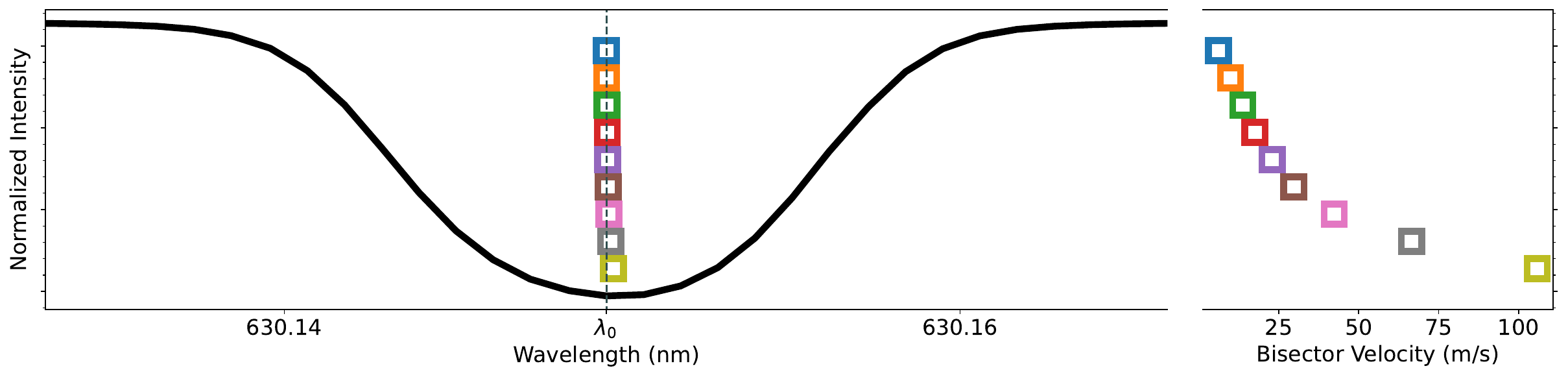}
    \caption{The Fe \RomanNumeralCaps{1} 630.1 nm line profile at the maximum electron beam heating rate time (Figure \ref{fig:Line_profiles} Model 1). Overplotted as colored squares at 10\% increments are the bisectors corresponding to the midpoint of the line profile at a given intensity. The dashed line corresponds to the rest wavelength, $\lambda_0$, of the transition. The right panel shows the same velocity bisectors converted to their Doppler velocity.}
    \label{fig:F12_Bisector}
\end{figure*}

The line contribution functions allow us to reconstruct the emergent intensities and recover the velocity information associated with the parts of the atmosphere that contribute to the line profiles. The line profiles shown for Model 1 in Figure \ref{fig:Line_profiles} exhibit a less dynamic evolution compared to the other models, remaining a relatively simple absorption profile throughout the course of the simulation. While Doppler shifts are not immediately apparent in the Model 1 flare-enhanced profiles, bisector analysis provides a useful diagnostic tool for detecting small-scale asymmetries. An example of this is shown in Figure \ref{fig:F12_Bisector} for the 630.1 nm line at the maximum beam heating time of Model 1 (Figure \ref{fig:Line_profiles}). In Figure \ref{fig:F12_Bisector}, the vertical dashed line shows the rest wavelength, and the square markers show the mid-point of the intensity profile at 10\% intervals between 10-90\% of the continuum level. The right panel shows the same bisectors converted to their Doppler velocities. With increasing depth into the line profile the markers shift progressively right of the dashed line, indicating m s$^{-1}$ scale redshifts, with greater redshift velocities closer to the line core. 

\begin{figure}[!h]
    \centering
    \includegraphics[width=\linewidth]{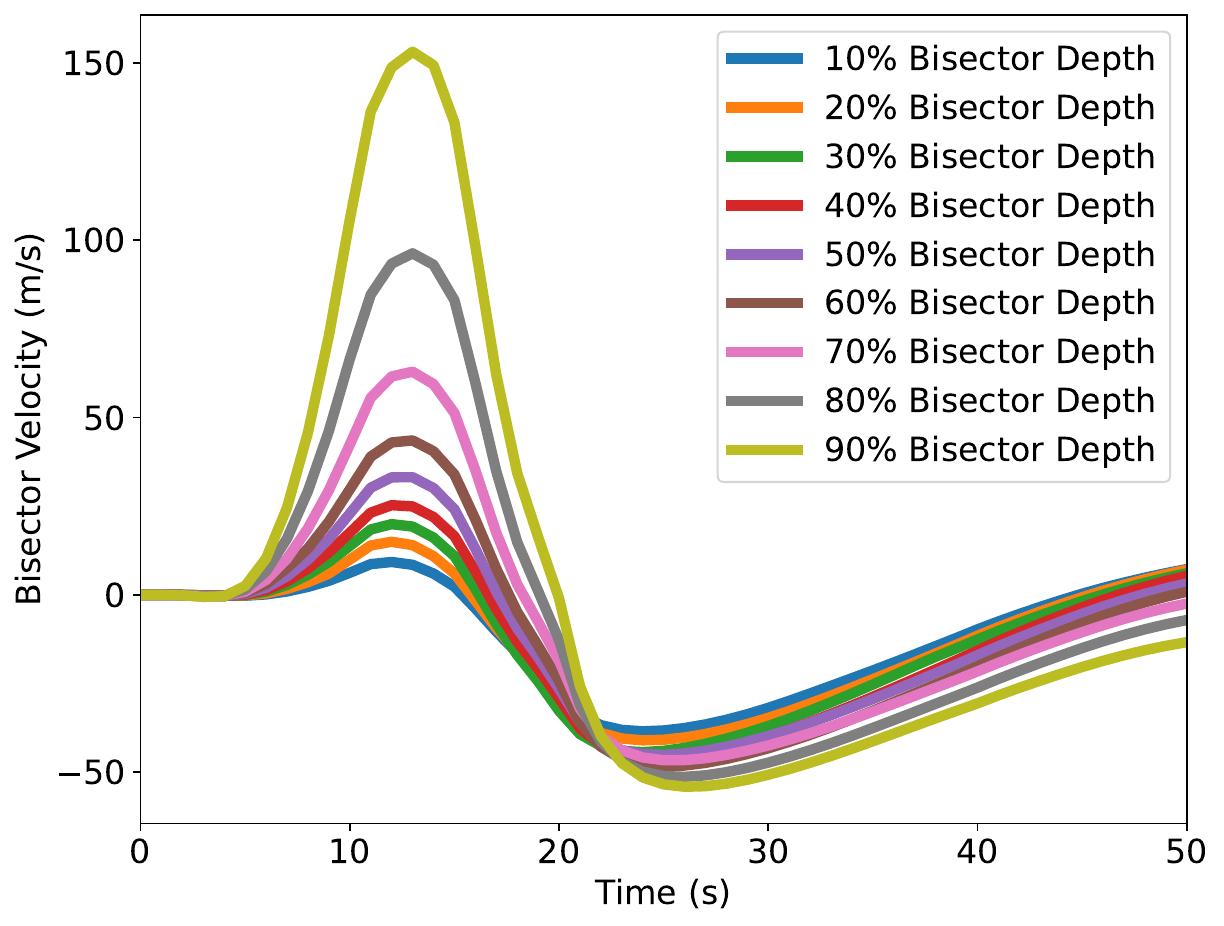}
    \caption{The temporal evolution of the 630.1 nm velocity bisectors throughout Model 1's duration. The color of each line corresponds to the square marker of the same color in Figure \ref{fig:F12_Bisector}. Positive velocity values indicate redshifts to the bisector, and negative values indicate blueshifts.}
    \label{fig:Bisectors_Time}
\end{figure}

This apparent redshift, indicating a downflowing photosphere, is only present during the periods of beam heating and evolves into a blueshift in all of the bisectors during the later stages of the model as shown in Figure \ref{fig:Bisectors_Time}. The colors of each line reflect the bisector depth of the same color in Figure \ref{fig:F12_Bisector}, showing how the Doppler shift of different components of the line profile evolves over the model. The cause of this switch in the Doppler shift was the primary investigation in Paper 1, where we refer the reader for a full description. In summary, the Fe I line profiles exhibited a misleading apparent redshift due to km s$^{-1}$ scale upflows in the lower chromosphere. Increased contribution to the Fe I lines from chromospheric heights resulted in a minor asymmetric intensity increase to the blue side of the line core, resulting in the total emergent profile appearing shifted to longer wavelengths. Once the beam heating ended, the contribution functions of these lines restored to being near-exclusively photospheric and the m s$^{-1}$ scale blueshifts caused by the upflows of the upper photosphere were accurate representations of the photospheric velocity field. The redshift to blueshift evolution of the Fe I line profiles shown in Paper 1 is also present in Model 1 (Figure \ref{fig:Bisectors_Time}). Throughout the evolution of the Model 1 line profiles, the extent of the Doppler shift is highly dependent on the bisector's depth into the line profile. This is especially the case during the initial apparent redshift where bisectors probing the line core indicate a ten times greater upflow velocity than the near-continuum bisectors. The maximum indicated velocity for all bisectors temporally coincides with the time of greatest enhancement to the line core ($<150$ m s$^{-1}$), several seconds after the "Maximum Heating" time in Figure \ref{fig:Line_profiles}, in agreement with Paper 1.

The complexities induced by the flare-enhanced chromospheric contribution are only further exhibited in  Models 2 \& 3 which show more dynamic line profiles (Figure \ref{fig:Line_profiles}). However, as shown in Figure \ref{fig:Density_Contrib} the chromospheric condensation plays a much more significant role in the contribution functions of the Fe I lines compared to Model 1. For Models 2 \& 3 we separate the contribution function profiles into three distinct regions based on the physical conditions of the atmosphere. By integrating the contribution on a height scale for a given region, the features observed in the emergent spectral line profiles may be isolated and matched temporally with the physical phenomenon in the RADYN model. For the purposes of this analysis we shall split the total emergent line profile shapes into individual contributions from three regions:

\begin{itemize}
    \item The photospheric component to the line profile that originates between the photospheric floor (0 km) and TMR. The amount of photospheric contribution remains largely constant throughout the flare both in Model 2 \& 3, with emission from regions higher in the atmosphere dictating the line profiles' shape evolution.

    \item The chromospheric component arises in the region between the temperature minimum and the point where the density increases at the start of the condensation.

    \item The condensation component, a downwards propagating region of increased density formed around the region of the largest direct beam heating, is driven by the expansion of MK material at greater atmospheric heights above. As the density and contribution functions drop orders of magnitude above the condensation, the upper chromospheric and coronal heights are also included in this classification but they have a negligible contribution to the profiles for the lines under consideration. 
\end{itemize}

\begin{figure*}[!h]
    \centering
    \includegraphics[width=0.9\linewidth]{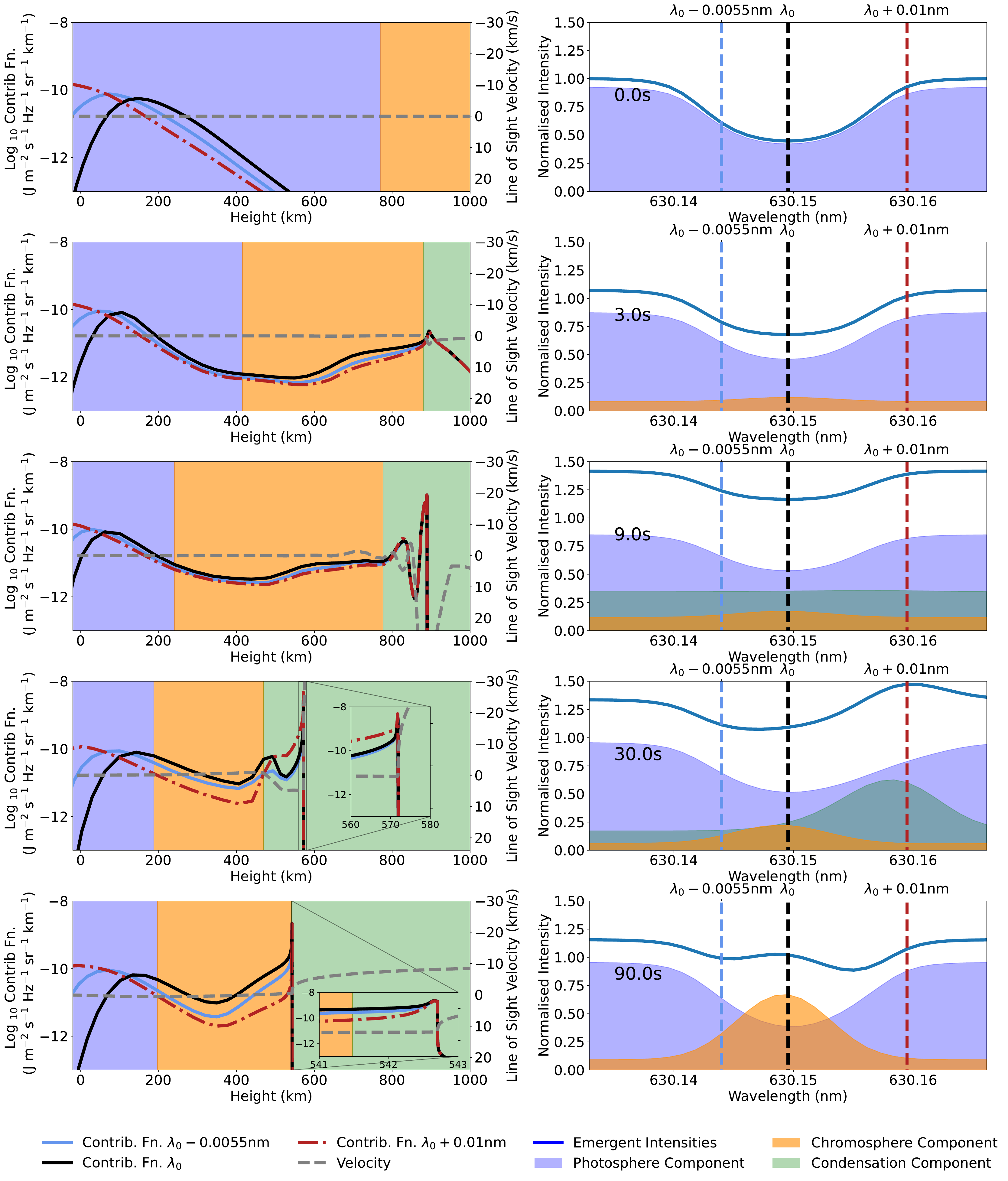}
    \caption{A breakdown of the 630.1 nm Fe \RomanNumeralCaps{1} spectral line into the three contributing regions of the lower solar atmosphere of Model 2. \textbf{Left Column:} Shaded zones refer to the classification of the height range in the atmosphere of the blue; photosphere, yellow; chromosphere and green; chromospheric condensation and above. The dashed grey line shows to the line-of-sight velocity in the RADYN model atmosphere at a given time, corresponding to the right axis. \textbf{Right column:} The emergent intensity profile of the 630.1 nm line for a stated time in the model and the integration of the contribution function in each zone Dashed colored lines show the wavelengths at which the left column height-scale contribution functions are in the line profile. Shaded zones below the emergent line profile indicate the fraction of the emergent intensity that is emitted from a given classification zone (photosphere, chromosphere, or condensation)}
    \label{fig:Zhu_Multipanel}
\end{figure*}

In Figure \ref{fig:Zhu_Multipanel} we show the three regions in the atmosphere of Model 2 and the contributions to the total line profile that emerges from each. The blue, yellow, and green shaded regions highlight the height ranges of the photosphere, chromosphere, and condensation regions respectively at the five stages of the model evolution discussed for Figures \ref{fig:Heating_profiles} \& \ref{fig:Line_profiles}. Over-plotted on the left column as colored lines are the contribution function profiles at three wavelengths of the 630.1 nm Fe \RomanNumeralCaps{1} line, sampling the formation of the blue wing (light blue solid line), line core (black line) and red wing (red dashed line). The wavelengths that the sample contribution functions represent are shown as dashed colored lines of the same colors in the right column plots of Figure \ref{fig:Zhu_Multipanel} at the appropriate wavelength.  Also shown in the left column plots is the line-of-sight velocity (dashed gray line). Positive values for the velocities correspond to downflows, and negative values to upflows. 

The right column of Figure \ref{fig:Zhu_Multipanel} shows the breakdown of the emergent line profiles into their constituent parts by integrating the contribution function over each of the shaded regions highlighted in the left column. 
The intensity profile of each component shows the emission from the height range of the same color in the left column plot. All values are normalized to the pre-flare continuum value as in Figure \ref{fig:Line_profiles}. The focus of this work is not to predict exact intensity values but to understand how profile shapes are formed during flare cases similar to  these models. It should be noted that at times there is a slight discrepancy between the total emergent profile and the height-integrated contribution functions, such as in the unshaded regions of the top right 0s pre-flare profile. This arises from inaccuracies in the numerical integration scheme over the irregularly spaced gridpoints utilized in the model. The calculated intensities between the two radiative transfer solutions are in good agreement to within a few percent at all times.

For Model 2, the pre-flare panel (top row; Figure \ref{fig:Zhu_Multipanel}) intensity profile is contributed near-exclusively from the photosphere. The contribution functions of the three wavelengths across the line profile show their maximum in the lower photosphere around 200 km or lower and drop by several orders of magnitude to insignificant emission from the chromosphere or above. There is no pre-flare line-of-sight velocity (dashed gray line) hence a symmetric line profile. During the initial heating (3s; second-row Figure \ref{fig:Zhu_Multipanel}) several changes in the atmospheric structure and contribution function are evident. Most notably, the beam heating reduces the photosphere by $\sim$350 km as the TMR region is heated.

The contribution functions from the chromosphere now show a significant increase at all three wavelengths, most significantly in the line core (black line, yellow-shaded region). However, all three contribution function lines in the left panel show similar profiles meaning that no particular part of the emergent line profile has a significant contribution from the chromosphere compared to other wavelengths. A secondary peak in the contribution functions is shown at $\sim900$ km co-spatial with the start of the formation of the condensation, noted here as the slight dip in the line-of-sight velocity of $-1$-$2$ km s$^{-1}$, where the density increase starts to form (Figure \ref{fig:Density_Contrib}). The deconstructed line profile in the right panel now shows that while the emergent intensity is still dominated by emission from the photosphere, the primary cause for the enhancement of the line core is from the increased chromospheric contribution which contributes $10-15$\% of the total emergent intensity. The depth of the line core slightly reduces as contribution from the chromosphere is slightly greater at these wavelengths than near the continuum.


The relative contribution of the chromosphere continues to increase throughout the evolution of the atmosphere. By the time of maximum heating (9s; Middle row panels) the TMR has decreased by another 200 km, further compacting the photospheric height range. The three contribution functions are largely similar to the Initial Heating time values throughout the photosphere and chromosphere, though the lower chromosphere shows an order of magnitude greater values. Again, wavelengths closer to the rest wavelength show slightly higher contribution function values. The most notable change however is the more developed condensation, noted by the dip of the line-of-sight velocity to -60 km s$^{-1}$ and the beginning of the green shaded region in the left column plots, and a lower slower-moving feature at $\sim 800$ km. By this time the condensation has swept up enough material to have a significant density increase compared to the pre-flare state and is still subject to the majority of the incident beam heating (Figure \ref{fig:Density_Bheat}).

All three contribution function lines show significant, order of magnitude, increases at the leading edge of these mass flows. While in the lower regions of the atmosphere the contribution functions of the three wavelengths were similar but not identical, at these early times of the condensation's evolution they show directly overlapping profiles. The similarity of the contribution functions within the condensation leads to the uniform intensity contribution shown in the right column emergent intensity breakdown, with the entire profile being equally contributed to by the green-shaded region. This is the primary cause of the entire line profile's intensity jump between the Initial Heating and Maximum Heating timesteps shown in Figure \ref{fig:Line_profiles}, enhancing the entire profile above the pre-flare continuum value. Meanwhile, the line core experiences increased filling-in, reducing the core depth, due to the contribution from the chromosphere which has slightly increased due to the contribution functions increase in the lower chromosphere. The photosphere continues to demonstrate a symmetric absorption profile (blue-shaded region) as in earlier timesteps, but is less dominant compared to earlier times in the model due to the increased emission from other regions.  

As the Model 2 atmosphere develops towards the Gradual Phase timestep, the fourth row, left column plot in Figure \ref{fig:Zhu_Multipanel} shows the condensation's further propagation through the lower solar atmosphere reaching a height of $\sim500$ km. By this time the condensation has considerably slowed down to $-5$ km  s$^{-1}$ due to the increased densities of the lower atmosphere. Additionally, the beam heating rate has significantly diminished during these later stages of the model (Figure \ref{fig:Heating_profiles}), reducing the localized energy input (Figure \ref{fig:Density_Bheat}) which previously sustained the condensation's momentum. This evolved state of the model demonstrates a clearer disparity for the three contribution function profiles across the lower solar atmosphere, where at previous times they had been relatively homogeneous showing similar enhancements regardless of wavelength. The line core and blue-side contribution functions (black and blue solid lines respectively) show enhancements throughout the lower chromosphere and at the leading edge of the condensation where the density is slightly increased but not yet swept up by the downflowing region. Conversely, the region of downward propagating material is coincident with the sudden rise of the dash-dotted red line, representative of the contribution function for the red-side wavelength, while the shorter wavelength contribution functions show slight reductions in their values. Additionally, the lower-photosphere maxima for the line core and blue-side contribution functions have restored to similar heights as in the pre-flare panel, while during the Initial Heating and Maximum Heating timesteps they had been $\sim 50$ km lower. Overall we see that longer wavelengths are preferentially formed within the downwards propagating condensation, though there is still some contribution to shorter wavelengths, and the lower atmosphere is more significant for wavelengths from the line core to the blue wing.

This is reflected in the deconstructed line profile for this timestep shown in the right column. The transient reshifted feature previously seen in Figure \ref{fig:Line_profiles} is primarily contributed from the condensation region shown by the green-shaded emission region at longer wavelengths.  The photosphere remains the primary contributing region to the total profile, defining the overall absorption profile of the line with intensity values very similar to those of the pre-flare panel in the top row. However, the blue-shaded region at this time shows an extended red-wing slope compared to the symmetric photospheric profiles at earlier times due to the re-absorption of photosphere-emitted photons in the leading edge of the condensation and upper chromosphere, while the blue side photons experience lower opacities from the red-shifted material. The chromospheric contribution component continues to primarily enhance the line core but shows a greater blueshift to its profile due to the $1-2$ km s$^{-1}$ upflows that have developed within the lower chromosphere.

While the photospheric contribution alone is similar to the pre-flare profile, the ``observed'' line profile shape is dominated by the influence of the chromospheric and condensation contributions. The enhancement of the total profile above the pre-flare continuum value arises largely due to the condensation's emission at near-continuum wavelengths, in particular at longer wavelengths which create the transient emission feature within the profile. A significant proportion of the emission in shorter wavelengths (the flat $\sim 20\%$ normalized intensity green-shaded region to the left side of the deconstructed line profile) arises from bound-free continuum emission which uniformly contributes across the entire line profile. The condensation's continuum emission is primarily emitted from the trailing edge of the downflow. This region can be seen as an inset in the left column panel showing the order of magnitude increase in all three contribution function lines over a very small $< 1$ km region. While the entire profile experiences this continuum enhancement to its intensities, the longer wavelengths show increased redshifted line contribution from the relatively dense unheated ($10^3$ K) front of the condensation. The line core (black solid line) and longer wavelength (red dashed-dot line) contribution functions in the left column panel show significant increases in emission at the leading edge and throughout the condensation, respectively. The optical opacity for the entire line profile above the condensation is extremely low due to the MK temperatures and orders of magnitude density reduction (Figure \ref{fig:Density_Contrib}).This causes the contribution function to dramatically decrease by orders of magnitude above the condensation, where insignificant amounts of emission to the line profiles are contributed.

By the late stages of Model 2, Figure \ref{fig:Zhu_Multipanel} bottom row, the lower atmosphere has begun to relax. The condensation feature, which had previously been downward propagating, has dissipated its momentum and the chromosphere begins to restore upwards.  The line-of-sight velocity structure is largely negligible in the lower solar atmosphere apart from a slow $<500 m s^{-1}$ upflow at the chromosphere-condensation boundary. The beginning of the condensation is still defined by the increase in density which was previously used to denote the start of the downwards propagating material, but unlike earlier analyzed timesteps this now denotes the mass front for the atmosphere restoring upwards once again. The contribution function continues to show an enhancement in the upper chromosphere within the region of mass density increase but at a reduced magnitude further away from the line core. The condensation contribution has shrunk to a $<2$ km region, shown by the green-shaded region highlighted in the inset zoom-in. As the condensation's contribution to emergent intensity is the integral of the contribution function over height, the small height scale indicates an insignificant contribution to the emergent line profile. This is indeed reflected in the emergent line profile breakdown shown in the right column. This shows a two-component contribution consisting of a stationary photosphere absorption component and the blue-shifted chromospheric material. The asymmetric central reversal is largely the product of the upflowing chromospheric plasma filling the line core to the blue side similar to Paper 1's asymmetry and the bisector velocities discussed for Model 1, but with a far greater chromospheric contribution due to the mass deposited by the condensation. This would indicate that photospheric lines are not just subject to chromospheric conditions during periods of significant heating, but also at later times if the mass drive from chromospheric condensations is sufficiently large.

\begin{figure*}[!h]
    \centering
    \includegraphics[width=0.9\linewidth]{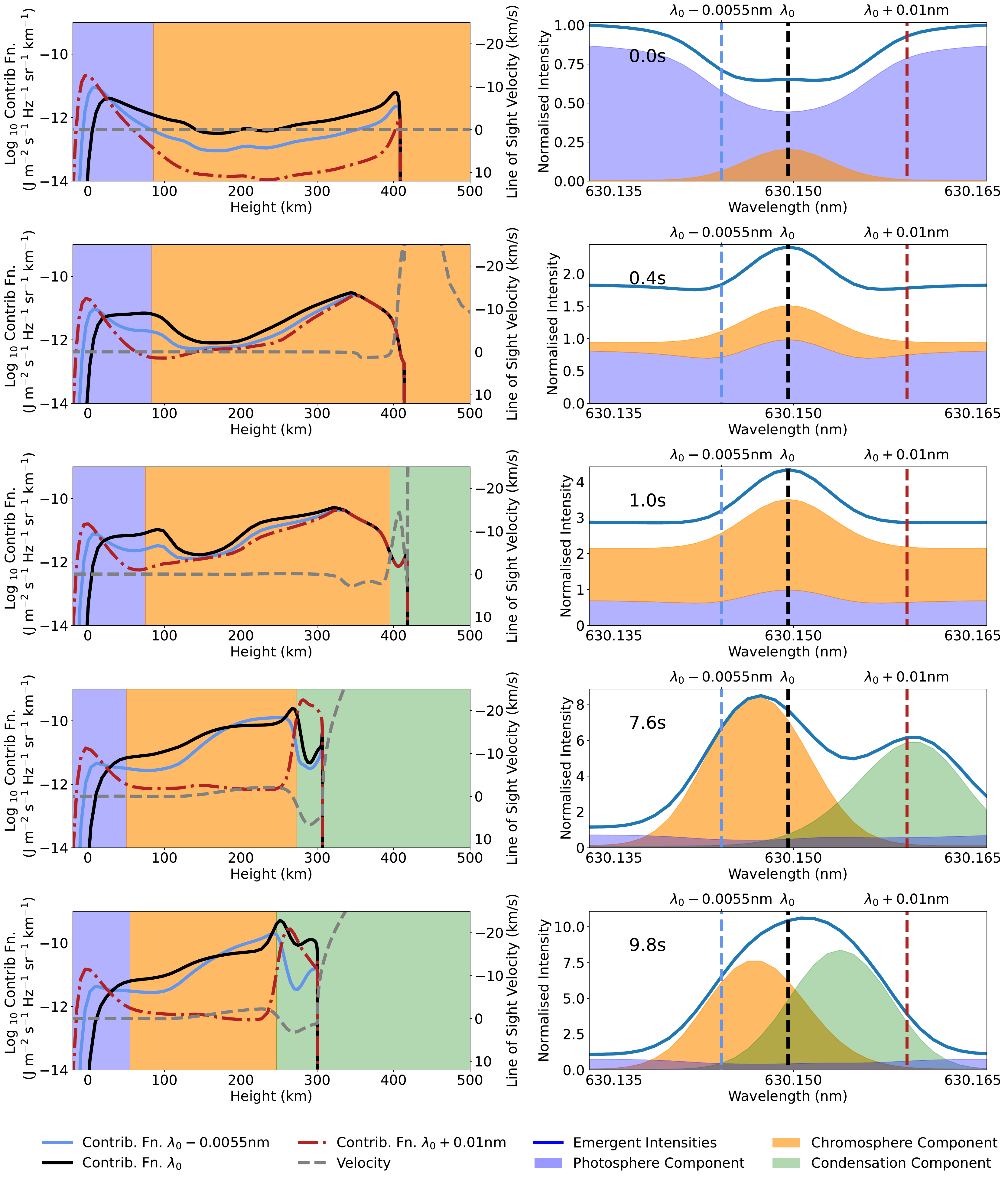}
    \caption{A breakdown of the 630.1 nm Fe \RomanNumeralCaps{1} spectral line into the three contributing regions of the dMe flare lower atmosphere (Model 3). All shaded regions and lines have the same meanings as in Figure \ref{fig:Zhu_Multipanel}. }
    \label{fig:Mdwarf_Multi}
\end{figure*}

A similar line profile reconstruction was conducted for the Model 3 stellar flare scenario and is shown in Figure \ref{fig:Mdwarf_Multi}. The pre-flare line profile (top right panel) differs from the solar models by the inclusion of a chromospheric component, not present in Model 2, contributing approximately 20\% of the line core intensity. The plateaued contribution function profiles in the cooler chromosphere in the M dwarf, along with the contribution function uptick at the base of the transition region at $\sim 400$ km. result in the central reversal profiles previously mentioned in Figure \ref{fig:Line_profiles}. During the initial heating, $<1$s, we see a much greater increase in the chromospheric contribution functions compared to Model 2. While all three wavelengths show an increase throughout the chromosphere, the increased contribution most significantly originates from the base of the transition region (300-400 km) below the 10's km s$^{-1}$ upflows of the evaporation front (second row, left column). 

The rise in contribution from this region is due to the larger mass density encountered by the electron beam compared to the solar atmosphere Models 1 \& 2 (Figure \ref{fig:atomos_profiles}). The high degree of localized beam heating given the short ramp-up timescale to maximum heating, 1s, discussed in Figure \ref{fig:Density_Bheat} promotes greater emission at low optical depths. The contribution from the upper chromosphere is sufficiently enhanced and contributes more to the line profile than the original photospheric component. This is shown in the right panel for this time (0.4s) as the yellow-shaded chromospheric component exceeds the blue-shaded fraction of the emergent intensity profile at all wavelengths. In the M-dwarf flare atmosphere, the line is chromosphere-dominated throughout the heating. The near-immediate absorption to emission switch of the line profile shape is a product of the chromospheric component and the centrally reversed photospheric component. The line core contribution enhancement in the photosphere close to the temperature minimum boundary (black line at $\sim75$ km) is a result of radiative backwarming, largely by Balmer continuum absorption, which heats the upper photosphere by $\sim200$ K.  
Given the lack of line-of-sight velocity in the lower atmosphere at this time in the model, a single peak enhancement around the rest wavelength from both components constitutes the total observed emission profile (solid blue line right panel). 

The single emission peak persists for several seconds through the time of maximum beam heating (1s; shown in the middle panels of Figure \ref{fig:Mdwarf_Multi}) with the relative strength of the emission peak varying depending on the chromospheric contribution. The fraction of the emission originating from the photosphere will remain semi-constant throughout the rest of the atmosphere's evolution. Wavelengths closer to the rest wavelength receive a slightly greater enhancement from the heating of the TMR region, shown by the decreasing plateauing of the contribution functions at the photosphere to chromosphere boundary. This lower region of the atmosphere exhibits negligible line-of-sight velocities at this time, so a greater contribution from the line core is to be expected with a lack of Doppler shifts. The beginning stages of the condensation are also shown around the period of maximum beam heating as the green-shaded region which was not present in earlier panels in the left column. By this time the initial density rise denoting the condensation has formed but has not swept up enough material to be a strong contributor and so all three wavelengths' contribution functions have relatively low values within the condensation compared to lower in the atmosphere.

As discussed for Figure \ref{fig:Line_profiles}, the emission peak exhibits a gradually increasing redshift during the following seconds after the maximum beam heating. While the entire profile's intensity drops with the continuum reduction, it still continues to be dominated by the chromospheric material at the base of the transition region. This is shown in Figure \ref{fig:Mdwarf_Multi}'s fourth row left column, where the line core and blue side contribution functions (black and light blue solid lines) are greatly enhanced throughout the entire upper chromosphere. At this time the panel indicates that shorter wavelengths are probing the km s$^{-1}$ upflows in the 200-300 km range in the chromosphere before the condensation reaches this region.

Throughout the several seconds between rows three and four, the condensation has accumulated material as it propagates downwards, increasing its contribution compared to earlier times. While the contributions for the line core and blue wing show nearly two orders of magnitude greater contribution in the chromosphere, compared to the red wing (yellow line), the reverse occurs in the condensation. The contribution functions of wavelengths blue of the line core rapidly drop off, the downwards red shifting material being unfavorable for their formation, as the red-side peaks in the $<10$ km s$^{-1}$ compression. The emergent profile shown in the right column clearly shows the source of the bimodal shape previously seen. A chromospheric-dominated blue lobe and a clear condensation-originating component to the red side, with very little meaningful contribution from the condensation to the blue side due to its opposing line-of-sight velocity to create blue-shifted photons.

By the End Model time 2s later (bottom panels), though the condensation has only propagated $\sim 20$ km lower into the chromosphere the $10^{-9}$ g cm$^3$ densities of the dMe chromosphere have significantly slowed down the feature. While the contribution functions in the left panel remain largely similar to the Gradual Phase time, the re-categorization of the swept-up chromospheric material by the condensation leads to a reduction in the chromospheric fraction of the emergent intensity.  With the slower downflow velocity now shown in the left panel making the condensation exhibit a lesser degree of red-shifting, the singular, broad emission spike previously described can be seen to be a product of both near equally contributing components. The photospheric component (blue-shaded region) is shown to only minimally and nearly uniformly contribute across the line profile but is the dominant contributor to the near continuum wavelengths.

In all, the line profile is formed by contributions from three distinct regions, each one determining different properties of the emergent profile shape. The photosphere generally maintains an absorption profile with intensity values similar to the pre-flare state of each model and is the primary contributor to near-continuum wavelengths. The chromospheric emission is relevant for the shorter wavelengths of the line core and the chromospheric condensation contribution creates transient redshifted features which affect the red wing intensities. A truly complex, multi-components line profile for what might traditionally have been expected to be a line formed through photospheric heating only.
 
\section{Conclusions}\label{sec:Conclusions}

We have utilized radiative-hydrodynamic models of solar and stellar flares to synthesize temporally evolving intensity line profiles of deep-forming Fe \RomanNumeralCaps{1} spectral lines. The contribution functions have allowed us to deconstruct complex line profiles into the major contributing features in the solar atmosphere and investigate the primary cause of spectral asymmetries and transient features. We have shown that what may typically be assumed as a photospheric diagnostic can be subject to atmospheric properties across the entire atmosphere, especially in extreme solar and stellar flare scenarios. 

By utilizing differing solar \& stellar atmospheric structures and unique electron beam heating profiles in each case, we have explored a range of parameters to discern the differing responses of the Fe \RomanNumeralCaps{1} 630.1 nm spectral line (Figure \ref{fig:Line_profiles}). In all cases, the emergent intensity profile was found to be a composite of emissions from multiple regions of the solar atmosphere. During the early stages of all models, from initial heating through to shortly after the maximum heating rate time (Figure \ref{fig:Heating_profiles}), increased contribution from the chromosphere was the primary cause for the increase in intensity across the entire line profile. In particular, the line core exhibited the greatest chromospheric ``filling in'' for the two considered solar cases (Models 1 \& 2).

For the quiet sun pre-flare atmosphere structure (Model 1) the total line profile exhibits a subtle (m s$^{-1}$ scale) redshift throughout the period of beam heating ($\leq$20s) followed by a developing blueshift at later times (Figures \ref{fig:F12_Bisector} \& \ref{fig:Bisectors_Time}). This presented the simplest two-component emission profile where emission from regions above the photosphere can result in falsely attributed Doppler information during a flare, in agreement with previous works \citep{Sharykin17,Hong18,Monson21}. Model 2 in this study, the cooler plage solar atmosphere, shows a greater enhancement in the line core compared to the quiet sun case Model 1, showing the same central emission feature asymmetry as shown for the 617.3 nm Fe \RomanNumeralCaps{1} line in \cite{Hong18} (Figure \ref{fig:Line_profiles}).

During the late stages of Models 2 \& 3, the chromospheric condensation formed as a result of extreme, localized beam heating in the upper chromosphere, becomes a dominant factor in the red wings. Due to the extreme $10^{13}$ ergs cm$^{-2}$ heating injected into Model 2, and the short timescale of the $10^{12}$ ergs cm$^{-2}$ dMe stellar flare heating Model 3 (10s), these models provide the necessary $10^{11}$ erg cm$^{-2}$ s$^{-1}$ electron beam flux requirements to drive explosive condensation flows \citep{Fisher85,Kowalski17}. The significant mass these features accumulate through their propagation, and the high degree of heating by the electron beam localized upon the condensation (Figure \ref{fig:Density_Bheat}), produce secondary peaks in the contribution functions of Fe I well above the photosphere (Figure \ref{fig:Density_Contrib}). 

This is shown to be a significant or dominant contributor to longer wavelengths of the spectral line, seen as an asymmetric brightness increases in the Gradual phase and Model End stages of Figure \ref{fig:Line_profiles} (middle row) and the secondary emission peak of Model 3's Gradual Phase. Model 2 has a sufficiently long development after the peak beam heating to show blue asymmetries, indicative that the 630.1 nm line is sensitive to the atmospheric restoration of the excess deposited material upwards again.

\cite{Graham20} found that the ``satellite'' red components of Fe \RomanNumeralCaps{2} 281.445 nm evolved three times too quickly compared to observations of the SOL2014-09-10T17:45 X-class flare, transiting the profile and merging with the rest wavelength peak. In addition, the relative emission of the satellite peak was found to be far too bright compared to the rest wavelength-based intensity peak, indicating a failure in the models to exaggerate this transiting component's emission. It has been suggested that the addition of turbulent flows in the condensation would Doppler broaden the satellite component, reducing the maximum intensity. 

In particular, the extreme spikes in the contribution functions at the chromosphere-condensation boundary noted for Model 2 (Figure \ref{fig:Zhu_Multipanel}) may prove to be unphysical. The one-dimensional limit of these models constrains mass motion to the magnetic field line direction, meaning the Doppler-shifted contribution spike may be exaggerated compared to the realistic three-dimensional motion of the plasma. This would lower the relative intensities of these satellite components in models closer to observations as momentum is dispersed out of the line-of-sight plane. If such consideration were applicable to the Fe \RomanNumeralCaps{1} lines to correct the relative intensity contribution from the red components in this work, it may reduce their contribution to a fraction of what is calculated for Models 2 \& 3. Regardless, it would still be necessary to evaluate these deep-forming spectral lines with a three-component framework, considering the evolution of the photosphere, chromosphere, and chromospheric condensations, to fully interpret the shape of the line profiles. 

With the advent of multi-dimensional flare models \citep{Cheung19,Druett23a,Druett23b} and the novel techniques created to expand the potential of 1D RHD models \citep{Osborne22}, the effects of this non-photospheric emission should be further investigated to detail the retrievability of accurate photospheric velocities during flares. 1D field-aligned RHD models are considered a reasonable simplification for including a particle beam as the primary energy input and atmospheric dynamics are field-aligned processes in the upper atmosphere \citep{Kerr20}. However, this simplification misses important 3D effects such as cross-field interactions that could limit the distribution of the beam energy  \citep{Allred23} or the accurate modeling of the heating induced by the changing coronal magnetic field and loop retraction \citep{Longcope2009,Cheung19}. Most relevant for the photospheric spectral lines studied in this work is how the 1D simplification differs from the 3D evolution of the lower atmosphere magnetic field.

For the quarter-circle loop structure of the model atmospheres used in this work, we are effectively sampling the spectral lines from a single kernel of the total loop arcade. However, the change in plasma beta in the lowest regions of the atmosphere can cause a deviation from the vertical, as the field lines are guided by the gas pressure. This may result in a dissociation between the field-aligned beam heating and radiative backwarming, the two primary heating mechanisms within the photosphere, that are not accurately treated in 1D. This may lead to a slight reduction in the heating of the lower atmosphere as the beam is redirected from the vertical and backwarming radiation must endure a longer path length to reach similar depths. We believe however that this effect, if present, would not result in significantly different flare-induced photospheric heating rates as the assumption of a near-vertical magnetic field within the photosphere is not unreasonable within active regions \citep{Wiegelmann14}.

Our work focuses mainly on the contributions from higher in the atmosphere, such as the chromospheric and condensation regions. In these regions, the primary missing feature inflicted by the 1D setup is the mixing of material from outside the flux tube which may affect the calculated temperatures, velocities, and densities. It is not unreasonable to assume that the plasma surrounding an individual flux tube is similarly heated by other loops within the arcade, so the effects of this mixing would be minimal with regions of similar height having a similar evolution to the surrounding plasma. Naturally, an observed spectrum will be a blend of this individual kernel and the surrounding plasma, and this is the primary limitation of 1D RHD models like RADYN. Multi-loop models \citep[][Mclaughlin et al. 2024, in preparation]{Kerr20,Rubio16} that attempt to recreate a 3D geometry by time-sequenced individual 1D models, have had mixed success. The development of these methods, alongside tools such as Lightweaver \citep{Osborne22} to simulate the unheated conjoining plasma between fluxtubes, presents the best opportunity to combine the NLTE effects, optically thick radiation treatment, and small spatial scales provided by 1D models with the realistic requirement of a 3D event.

\section{Acknowledgments}

The FCHROMA grid of flare models was produced under funding from the European Community’s Seventh Framework Programme (FP7/2007-2013) under grant agreement no. 606862 (F-CHROMA), and from the Research Council of Norway through the Programme for Supercomputing. We thank the anonymous referee for their comments which significantly improved the clarity of this paper. AJM acknowledges funding from the Science Technology Funding Council (STFC) grant ST/T506369/1. MM acknowledges support from STFC grants ST/P000304/1 \& ST/T00021X/1. AFK acknowledges funding from NSF Award 1916511.

\bibliography{paper}{}
\bibliographystyle{aasjournal}

\appendix   
\section{Additional Spectral Line Profiles}

\begin{figure}
\figurenum{10}
\plotone{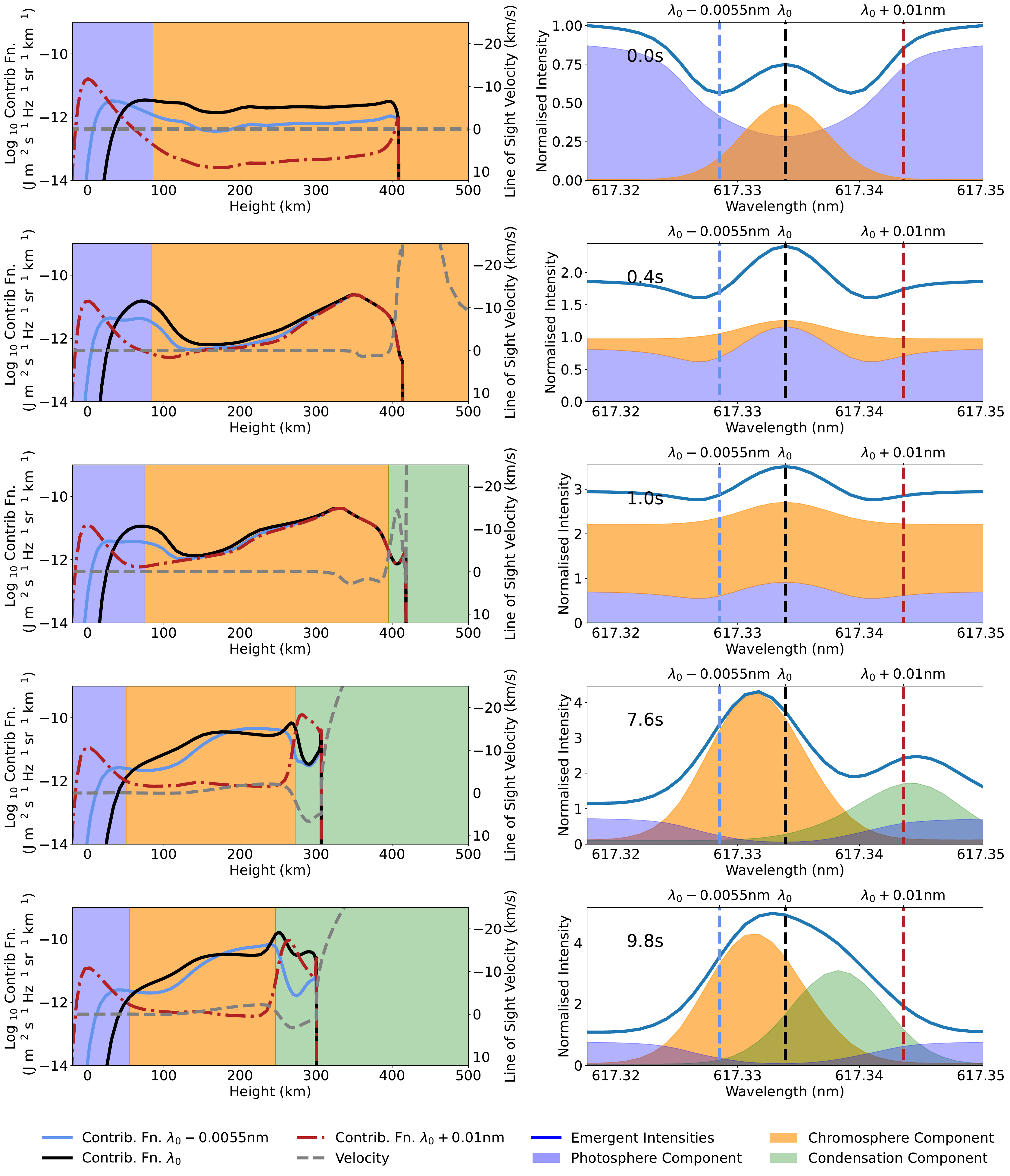}
\caption{The contribution function breakdown for the 617.3 nm Fe \RomanNumeralCaps{1} line from Model 3 at the same sampled times as in Figure \ref{fig:Mdwarf_Multi}. All lines and shaded regions have the same meaning as in Figure \ref{fig:Zhu_Multipanel}. The complete figure set of the 617.3 nm \& 630.2 nm spectral line breakdowns for Model 2 and Model 3 (4 images) is available in the online journal.}
\label{fig:Mdwarf_6173_Multi}
\end{figure}

The breakdown of the Fe \RomanNumeralCaps{1} 617.3 nm and 630.2 nm spectral line shapes for Models 2 \& 3 in the same manner as presented for Figures \ref{fig:Zhu_Multipanel} \& \ref{fig:Mdwarf_Multi} are shown in Figureset \ref{fig:Mdwarf_6173_Multi} available on the online journal. The Model 3 617.3 nm contribution function breakdown is sown in Figure \ref{fig:Mdwarf_6173_Multi} in the pdf version of the document. These two lines show very similar temporal evolutions to their contribution functions as discussed for the 630.1 nm line previously, but the fraction each atmospheric region contributes to the total emergent profile does differ. 

For Model 2, the two additional spectral lines show a weaker chromospheric contribution to the total profile, resulting in a weaker red asymmetry and central reversal to the line core at later times in the models compared to the 630.1 nm line shown in Figure \ref{fig:Zhu_Multipanel}. In Model 3 however the contribution from the condensation is slightly weaker for these additional lines, resulting in the less intense satellite feature in the line profile shape in Figure \ref{fig:Mdwarf_6173_Multi} (fourth row), and the more asymmetric singular emission peak (bottom row) compared to in Figure \ref{fig:Mdwarf_Multi}. The differing behavior of these three Fe \RomanNumeralCaps{1} transitions within the same model shows the importance of NLTE modeling of deep-forming spectral lines during flares to fully characterize their profiles' origins.

\figsetstart
\centering
\figsetnum{10}
\figsettitle{Additional Fe \RomanNumeralCaps{1} Spectral Line Contribution Functions Breakdowns}

\figsetgrpstart
\figsetgrpnum{10.1}
\figsetgrptitle{Model 2 617.3 nm Contribution Function Breakdown}
\figsetplot{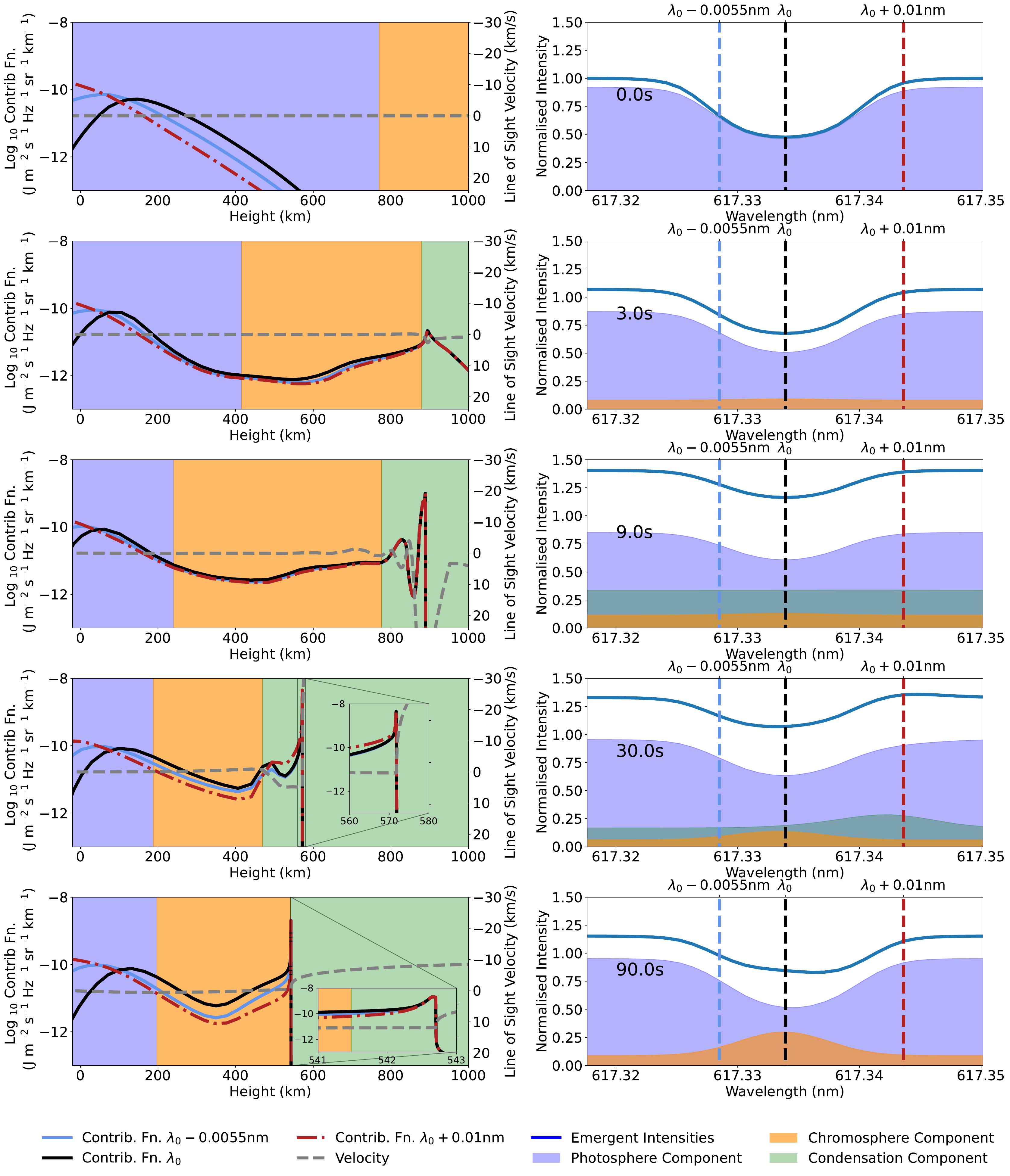}
\figsetgrpnote{The contribution function breakdown for the 617.3 nm Fe \RomanNumeralCaps{1} line from Model 2 at the same sampled times as in Figure \ref{fig:Zhu_Multipanel}. All lines and shaded regions have the same meaning as in Figure \ref{fig:Zhu_Multipanel}.}
\figsetgrpend

\figsetgrpstart
\figsetgrpnum{10.2}
\figsetgrptitle{Model 2 630.2 nm Contribution Function Breakdown}
\figsetplot{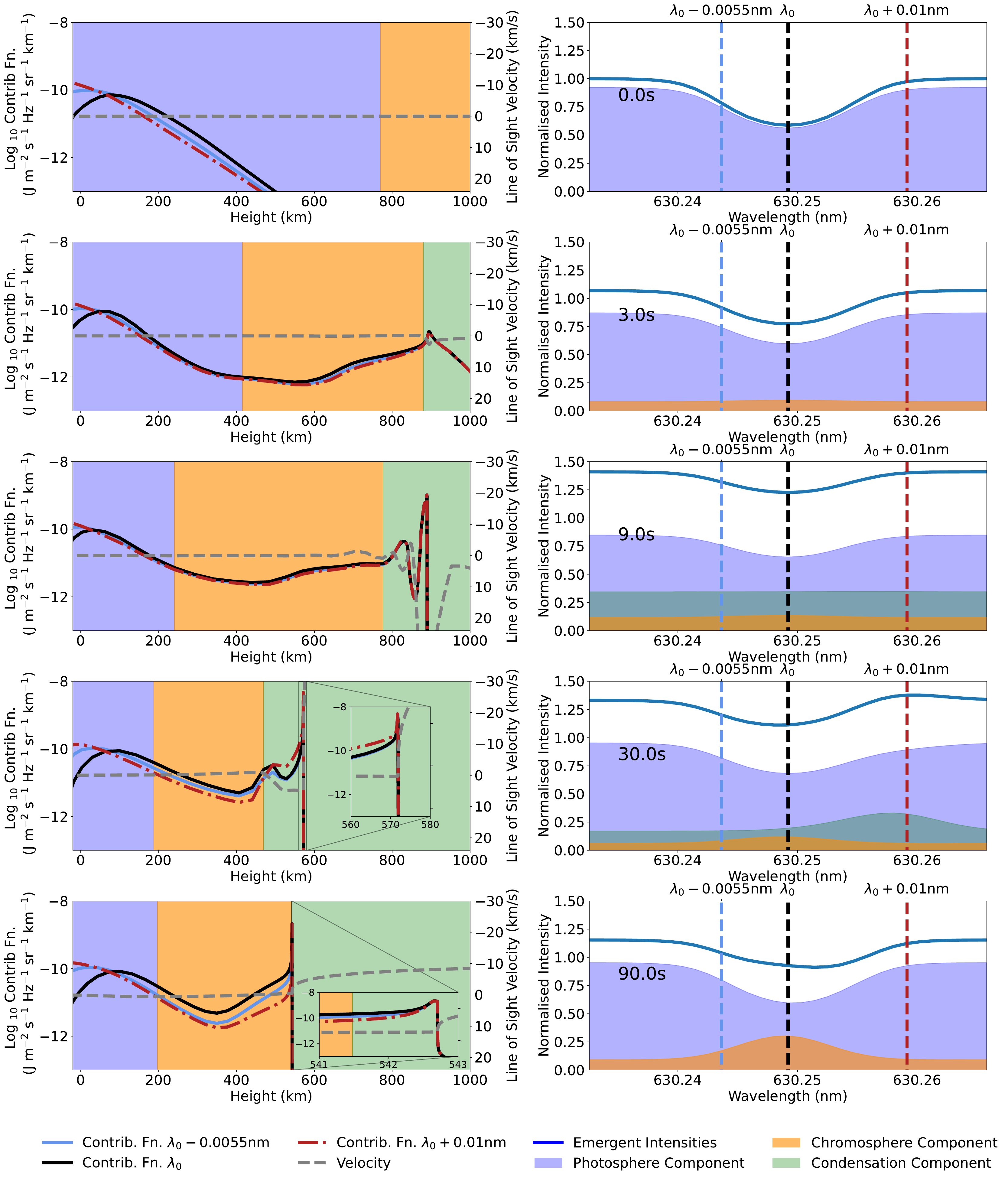}
\figsetgrpnote{The contribution function breakdown for the 630.2 nm Fe \RomanNumeralCaps{1} line from Model 2 at the same sampled times as in Figure \ref{fig:Zhu_Multipanel}. All lines and shaded regions have the same meaning as in Figure \ref{fig:Zhu_Multipanel}.}
\figsetgrpend

\figsetgrpstart
\figsetgrpnum{10.3}
\figsetgrptitle{Model 3 617.3 nm Contribution Function Breakdown}
\figsetplot{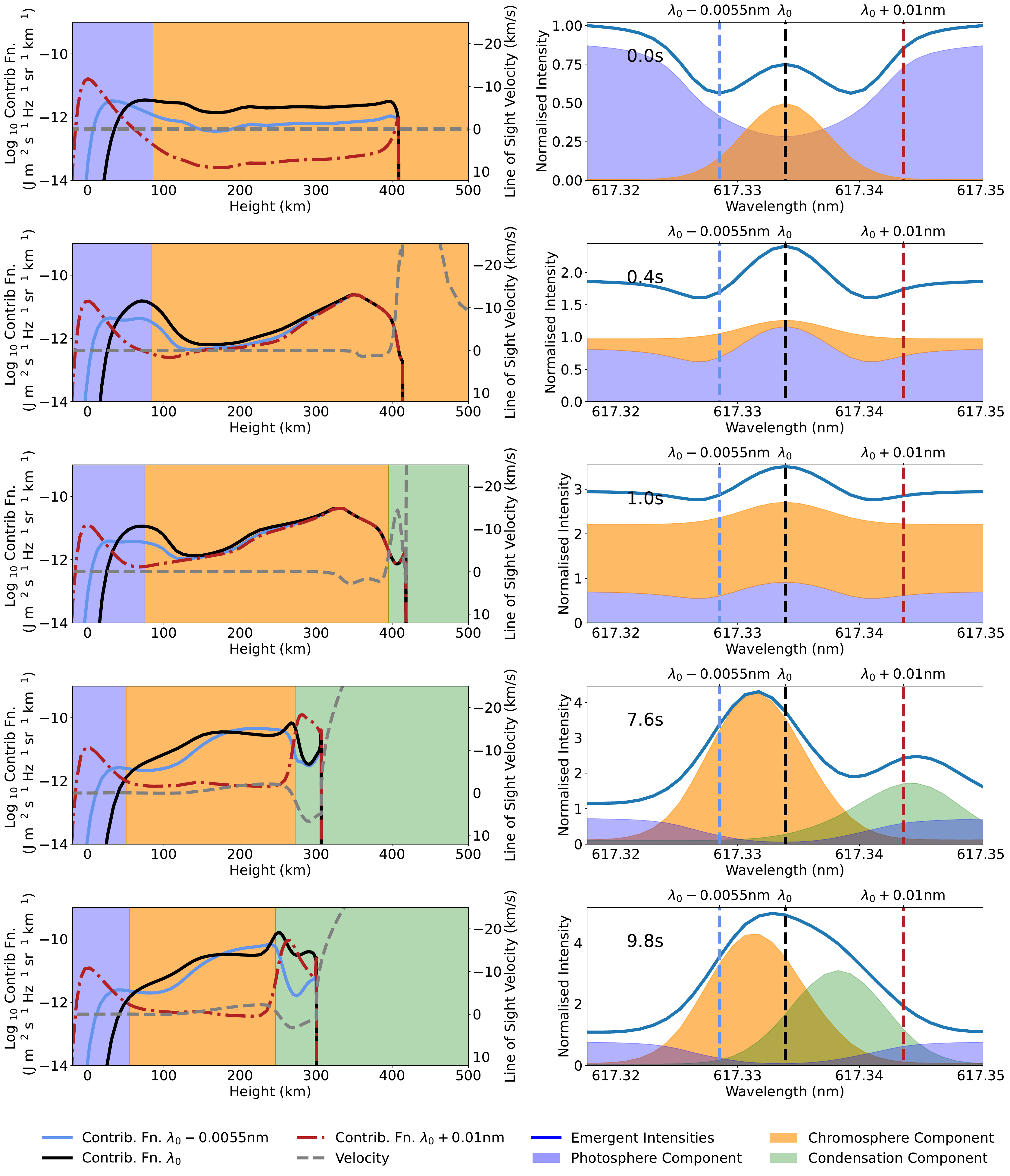}
\figsetgrpnote{The contribution function breakdown for the 617.3 nm Fe \RomanNumeralCaps{1} line from Model 3 at the same sampled times as in Figure \ref{fig:Mdwarf_Multi}. All lines and shaded regions have the same meaning as in Figure \ref{fig:Zhu_Multipanel}.}
\figsetgrpend

\figsetgrpstart
\figsetgrpnum{10.4}
\figsetgrptitle{Model 3 630.2 nm Contribution Function Breakdown}
\figsetplot{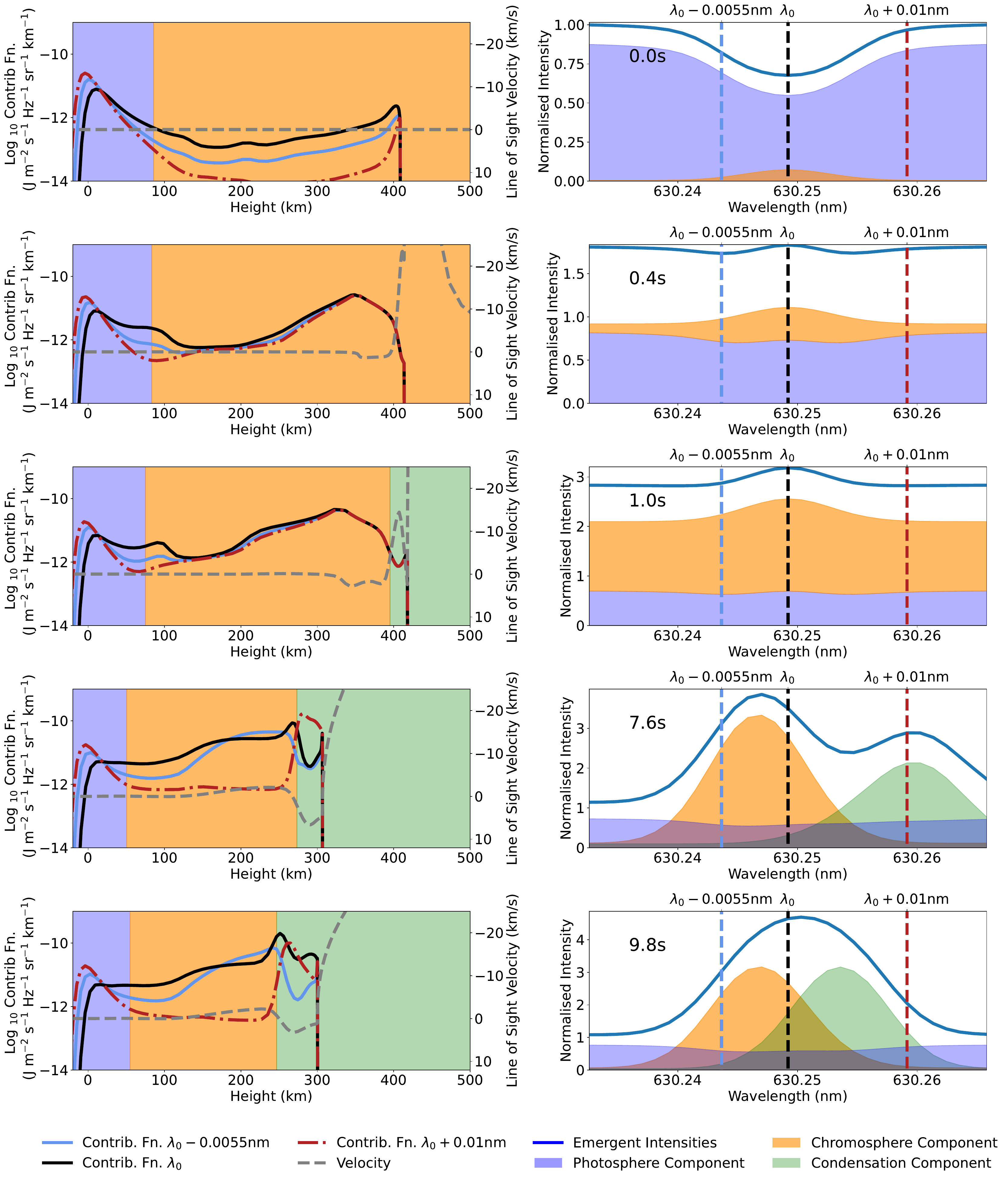}
\figsetgrpnote{The contribution function breakdown for the 630.2 nm Fe \RomanNumeralCaps{1} line from Model 3 at the same sampled times as in Figure \ref{fig:Mdwarf_Multi}. All lines and shaded regions have the same meaning as in Figure \ref{fig:Zhu_Multipanel}.}
\figsetgrpend
\label{figset:Contrib_Panels}
\figsetend

\end{document}